\documentclass[fleqn,10pt]{article}
\usepackage{amsmath,amsfonts,amssymb}
\usepackage[T1]{fontenc}
\usepackage[utf8]{inputenc}
\usepackage{ushort}
\usepackage{graphicx} 
\usepackage{caption}
\usepackage{subcaption}
\usepackage{xcolor}

\usepackage{bbold}
\usepackage{bbm}
\usepackage{bm}
\usepackage{array}
\usepackage{tikz}
\usepackage{tikz-cd}
\usepackage{scalerel}

\usetikzlibrary{decorations.markings}
\tikzset{
  negate/.style={
    decoration={
      markings,
      mark= at position 0.5 with {
        \node[transform shape] (tempnode) {$/$};
      },
    },
    postaction={decorate},
  },
}

\newcolumntype{L}{>{\centering\arraybackslash}m{3cm}}

\newcommand{\RN}[1]{%
  \textup{\uppercase\expandafter{\romannumeral#1}}%
}
\def\dunderline#1{\ushort{\ushort{#1}}}
\def\bs#1{\boldsymbol{#1}}
\def\bp{\begin{pmatrix}}
\def\ep{\end{pmatrix}}

\def\id{\ensuremath{\mathbbm{1}}}

\def\conj{\ensuremath{\mathrm{c}}}
\def\vecp{\ensuremath{\boldsymbol{\times}}}
\def\tr{\ensuremath{\mathrm{tr}}}

\def\beq{\begin{equation}}
\def\eeq{\end{equation}}
\def\bsubs{\begin{subequations}}
\def\esubs{\end{subequations}}
\def\bal#1\eal{\begin{align}#1\end{align}}

\newcommand{\mycdot}{{%
\mathchoice
{\cdot}
{\cdot}
{\cdot}
{\cdot}
}}

\makeatletter
\newcommand\etal{\emph{et al}\@ifnextchar.{}{.\@}}
\newcommand\etc{etc\@ifnextchar.{}{.\@}}
\newcommand\ie{i.e\@ifnextchar.{}{.\@}}
\newcommand\eg{e.g\@ifnextchar.{}{.\@}}
\makeatother

\begin{document}
\title{Nonlinear effects in Thomas precession due to the interplay of Lorentz contraction and Thomas--Wigner rotation}
\author{Antonio Di Lorenzo}
\date{Instituto de F\'{\i}sica, Universidade Federal de Uberl\^{a}ndia, Av. Jo\~{a}o Naves de \'{A}vila 2121, 
Uberl\^{a}ndia, Minas Gerais, 38400-902, Brazil}%
\maketitle
\begin{abstract}
It is demonstrated that the 3--vector $\bs{S}$ currently associated to the spin in an inertial frame does not contract, but rather dilates, in the direction of the velocity. The correct vector $\bs{T}$ is individuated. The equation of motion for the two vectors is shown to contain two terms, a common linear rotation, identified with Thomas precession, and also a nonlinear rotation depending on the direction of the spin itself.
\end{abstract}
\section*{Introduction}
Often, particles carry with themselves properties, \eg\ charge, mass, magnetic moment, electric dipole moment. 
While the motion of a particle $O'$ is described in a simpler form in an inertial reference frame, the evolution of physical quantities transported along $O'$ appears simpler when described in the comoving frame, even though it is an accelerated frame. Think, for instance, of a gyroscope attached to a car, which may accelerate, make a curve, brake, etc. Meanwhile, the car is being carried by Earth, which is spinning and rotating around the Sun, which in its turn is rotating around the center of the Galaxy. Even though the interior of the car is not an inertial reference frame, it is much easier to describe the motion of the gyroscope in such noninertial frame, after introducing noninertial forces that one attributes to gravitational fields, rather than describe its motion relative to the fixed stars. However, if from the car you look out of the window, all bodies not being transported along with you --- trees, mountains, the Sun, the planets, the stars in the night sky --- will appear to follow complicated, inexplicable motions, and there are no masses around compatible with the noninertial effects that you believed to be gravitational. If you want to simplify the description of the outside universe you should eventually switch to an inertial reference frame. 
The equations relating the change of the angular momentum of the gyroscope with the applied forces and torques should then be written back in an inertial frame, in which the movement of the origin of the gyroscope can be explained causally, \ie\ all accelerations can be associated to other bodies, which are usually limited to nearby bodies due to the short range of all forces except gravity.

In a given reference frame, physical quantities are described by scalars, three--dimensional vectors, and $3\times 3$ matrices. 
The special theory of relativity, despite its name, allows to describe the equations of physics in an absolute form, by making use of abstract mathematical objects in Minkowski space, which encompasses space and time. Such objects --- scalars, four--dimensional vectors, more generally tensors --- describe absolute quantities. One can then deduce the physical quantities in any reference frame as representations of these absolute objects. 

In our example, the angular momentum of the gyroscope, in the reference frame inside the car, is represented as a 3--vector $\boldsymbol{S}'$, 
which is 
a representation $\{0,S'^1,S'^2,S'^3\}$ of the abstract 4--vector $\underline{S}$ that is being transported along the car. In another reference frame, the same 4--vector is represented differently, as $\{S^0,S^1,S^2,S^3\}$. Here, the first component is the time component. 

\section*{Results}
\begin{figure}[h]
\begin{centering}
\includegraphics[width=0.99\textwidth]{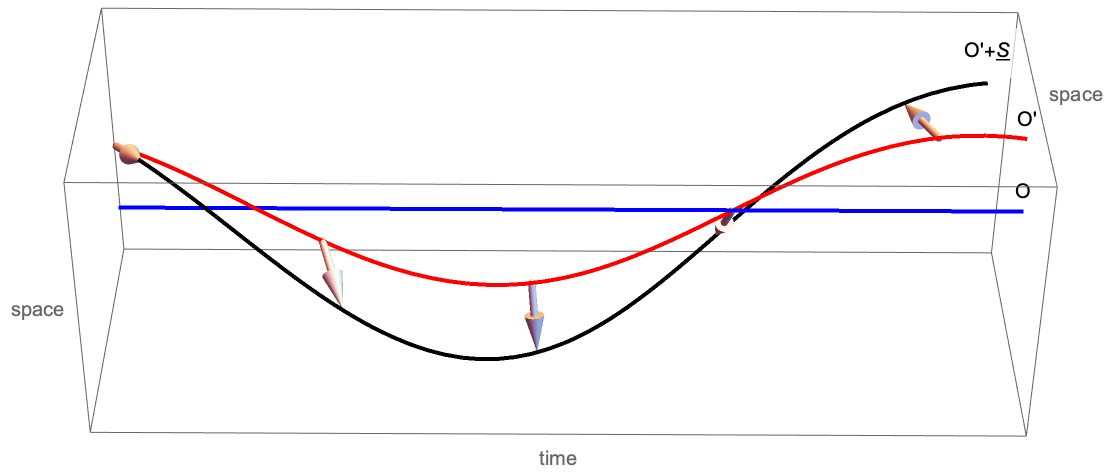}
\caption{\label{fig:trans} A vector field (arrows, sampled at discrete times) is being transported along an accelerated worldline $O'$ (red) as represented by an inertial observer $O$ (blue). Its tip traces a line (black) close to $O'$. The vector field is normal to the worldline $O'$ in the Minkowski metric, hence it does not appear as orthogonal in our Euclidean space representation.}
\end{centering}
\end{figure}

With reference to Fig.~\ref{fig:trans} we ask: given the 4--vector field $\ushort{S}_t$ transported along a worldline $O'$ in Minkowski space, normally to the world line, so that it is represented as a purely spatial vector, a 3--vector, in the comoving frame $\mathfrak{S}'$, how is it represented, as a 3--vector, 
in an inertial frame $\mathfrak{S}$? what are the equations of motion for the new representation, supposing the equations in the comoving frame to be known? 
These questions are particularly relevant in atomic physics, since an electron orbiting an atom carries an intrinsic magnetic moment. 

Our first result is that the relativity of simultaneity implies that the spin, as a 3--vector in the lab frame, 
is not the one which has appeared so far in the literature, consisting in the spatial part $\bs{S}$ of the 4--vector representation $\{S^0,\bs{S}\}$. Indeed, $\bs{S}$ does not undergo Lorentz contraction in the direction of motion, 
as a properly defined 3--vector should. On the contrary, it dilates in the direction of motion. 

The correct 3--vector to describe the spin in the lab frame, instead, 
is 
$\bs{T}=\bs{S}- S^0\bs{v}/c^2$. Because the spin is normal to the world line of the particle, in an inertial frame, $S^0=\bs{v}\mycdot\bs{S}$, 
while in a general frame $S^0=(\bs{v}-\bs{\Omega}\times\bs{r})\mycdot\bs{S}/[(1-\bs{g}\mycdot\bs{r})^2-(\bs{\Omega}\times\bs{r})^2+(\bs{\Omega}\times\bs{r})\mycdot\bs{v}]$, 
with $\bs{g}$ the equivalent gravitational field, $\bs{\Omega}$ the gravi--magnetic Coriolis field, and $\bs{r}$ the position of the spin. 
The combinations $\theta=1-\bs{g}\mycdot\bs{r}$ and $\bs{w}=\bs{\Omega}\times\bs{r}$ are to be identified with the scalar and the vector potential of the gravitational field. 

The previous literature uses $\bs{S}$ as the spin, 
and it is common to indicate as $\bs{V}$ the spatial components of a 4--vector $\vec{V}$, hence we are keeping the symbol $\bs{S}$ for the conventional spin for ease of comparison. We shall also write formulas for both the spin definitions. 
We shall see later on that, irrespectively of the procedure used, the linear rotation, seen in an inertial reference frame, 
of the 3--vectors $\bs{S}$ and $\bs{T}$ is the same. 

Our second result is that both vectors $\bs{S}$ and $\bs{T}$, in the lab frame, obey a differential equation which contains a linear rotation term and also a nonlinear term. The linear term, which we identify with Thomas precession, differs from the accepted value. We explain that the difference is due to the fact that the accepted value of Thomas precession does not describe the rotation of the spin in the lab frame, but rather is the rate of the Coriolis rotation of an auxiliary frame which is purely rotating relative to the lab. 

\subsection*{Detailed results}
Let us provide more details. 
According to a naive approach, which takes into account only the relativistic transformation of the electromagnetic field, the spin--orbit term in the Hamiltonian for the electron should be about twice as 
large as the one observed. 
A full relativistic treatment, however, reproduces the correct spin--orbit coupling, as shown by Thomas \cite{Thomas1926,Thomas1927}. 
The additional precession, due exclusively to relativistic effects for an accelerated frame, has been calculated in various ways, and there are some discrepancies in the literature. The common value accepted for the precession is \cite{Thomas1927,Jackson} 
\beq
\label{eq:wrong}
\boldsymbol{\omega}^\mathrm{Th} \stackrel{?}{=} -\frac{\gamma-1}{v^2} \boldsymbol{v}\vecp \boldsymbol{a}=-\frac{\gamma^2}{(\gamma +1)c^2} \boldsymbol{v}\vecp \boldsymbol{a} ,
\eeq
with $\bs{v}$ and $\bs{a}$ the Newtonian 3--velocity and 3--acceleration, $c$ the speed of light, 
and $\gamma=(1-v^2/c^2)^{-1/2}$ the Lorentz factor.
There are several other expressions for Thomas precession in the literature, some even with opposite signs, due mainly to some different interpretation or downright misinterpretation \cite{Malykin,Ritus}.
While the work of Malykin \cite{Malykin} reports exhaustively and critically the values that have been proposed, it ends up agreeing on the value as proposed by Thomas \cite{Thomas1927}.
In the following, we shall show that the accepted factor is incorrect, in the sense that Eq.~\eqref{eq:wrong} 
does not describe the precession of the spin as seen in an inertial frame, \eg\ the lab frame: 
because of Lorentz contraction, the spin of an accelerated particle makes a complicated motion, which is not a simple precession. 
Instead, we prove that Eq.~\eqref{eq:wrong} describes rather the rotation, relative to the lab frame, of the axes of a rotating frame (whose origin coincides with that of the lab frame). 
The relative rotation rate is thus a Coriolis rotation. 
This fact has been occasionally pointed out in the literature, \cite{Farago,Garg,Stepanov2012}. 

We show that, in the lab frame, the motion of the direction of the spin 
\beq
\frac{d}{dt}\bs{\Hat{S}}_t=[\boldsymbol{\omega}^\mathrm{Th}(t)+\boldsymbol{\omega}^\mathrm{nl}(t,\bs{\Hat{S}}_t)]\times \bs{\Hat{S}}_t
\eeq
contains a linear kinematic rotation, 
\beq\label{eq:right}
\boldsymbol{\omega}^\mathrm{Th} = -\frac{\gamma^2}{2c^2} \boldsymbol{v}\vecp \boldsymbol{a} ,
\eeq
but also an additional nonlinear precession, which, for a planar motion, is 
\beq\label{eq:right2}
\boldsymbol{\omega}^\mathrm{nl} = \frac{\gamma^2}{c^2} \left[\bs{v}\mycdot\bs{a}\, \bs{\Hat{v}}\mycdot\bs{\Hat{S}}\,\bs{\Hat{n}}\mycdot\bs{\Hat{S}}
+\frac{1}{2}\kappa v^3 \left(\bs{\Hat{n}}\mycdot\bs{\Hat{S}}\,\bs{\Hat{n}}\mycdot\bs{\Hat{S}}-\bs{\Hat{v}}\mycdot\bs{\Hat{S}}\,\bs{\Hat{v}}\mycdot\bs{\Hat{S}}\right)\right] \bs{\Hat{b}}.
\eeq
Here $\kappa$ is the curvature, $\bs{\Hat{v}}$ the tangent, $\bs{\Hat{n}}$ the normal, and $\bs{\Hat{b}}=\bs{\Hat{v}}\vecp\bs{\Hat{n}}$ the binormal of the trajectory described in the lab frame. 
The nonlinearity is a consequence of splitting the motion of the spin in an equation for its norm and an equation for its direction, as detailed in the Methods section. The overall equation for the spin vector, however, is linear, and it is but the Bargmann--Michel--Telegdi equation \cite{BMT} represented in the lab frame. 
The presence of a nonlinearity was found out by Stepanov \cite{Stepanov2012} in a particular case, and also by Kholmetskii and Yarman \cite{Kholmetskii2020}. 

Furthermore, we shall prove that for a uniform circular motion, for which the linear precession rate 
\eqref{eq:right} is constant, the average contribution of the nonlinear term \eqref{eq:right2}, when added to the linear term yields Eq.~\eqref{eq:wrong}. This apparently fortuitous occurrence may be at the origin of the unquestioned standing of Eq.~\eqref{eq:wrong}. 

Our Eq.~\eqref{eq:right} has a simple geometrical interpretation for planar motions: 
the angle of linear precession during a time $T$ is equal to the area swiped 
by the celerity $\gamma \bs{v}$ divided by $c^2$. The corresponding lowest order limit, with the factor $\gamma=1$, was mentioned by Borel \cite{Borel,Borel2} already in 1913. 
Thus, apparently, Borel had derived at least an approximate expression for Thomas precession, using merely  geometrical considerations.

In solving the problem of Thomas precession, we used a coordinate--free approach to the problem, which allows us to identify an skew--symmetric tensor $\dunderline{\mathfrak{B}}$, the generator of Fermi--Walker transport, depending only on the intrinsic geometry of the worldline of the accelerated particle, which describes a kinematic rotation in Minkowski space. The coordinate--free approach is extremely valuable, in that it is robust against error, and in that it provides a manifestly invariant picture, in the spirit of Minkowski \cite{Minkowski} and of the reference book by Misner, Thorne, and Wheeler \cite{MTW}. 
Often we read that the equations of physics must be covariant. Actually, covariance implies already the choice of a representation, where the physical quantities are represented by lists with upper or lower indexes, which transform contravariantly or covariantly. However vectors and tensors are invariant objects, only their representations are contravariant or covariant.

%
In the lab frame, the motion of the spin 
 is dictated by  
\bsubs
\label{eq:atomfull}
\bal
\partial_t \bs{S}_t =\ \frac{q}{m} \biggl\{&\frac{g_\mathrm{L}}{2\gamma_t}\left[\bs{S}_t\vecp\bs{B}_t+\bs{S}_t\mycdot\bs{v}_t \, \bs{E}_t\right]
-\left(\frac{g_\mathrm{L}}{2}-1\right)
\bs{S}_t\mycdot\bs{G}_t\, \bs{v}_t \biggr\}\ ,
\label{eq:atomfulla}
\\
\partial_t \bs{T}_t =\ \frac{q}{m}\biggl\{&\frac{g_\mathrm{L}}{2\gamma_t}\left[\bs{T}_t\vecp\bs{B}_t-\bs{T}_t\mycdot\bs{E}_t\,\bs{v}_t\right]
+\left(\frac{g_\mathrm{L}}{2}-1\right)
\bs{T}_t\mycdot\bs{v}_t\,\bs{G}_t\biggr\}\ .
\label{eq:atomfullb}
\eal
\esubs
Equation~\eqref{eq:atomfulla} is the spatial part of the Bargmann--Michel--Telegdi equation represented in the lab frame. 
Here, the term not multiplied by 
$g_\mathrm{L}$, $(q/m)\bs{S}_t\mycdot\bs{G}_t\, \bs{v}_t $, identifies the kinematic contribution due to the spin being accelerated; 
the other terms are due to the direct coupling with the electromagnetic field. 
The 3--vector $\bs{G}=\gamma\left[\bs{E}-\bs{E}\mycdot\bs{v}\,\frac{\bs{v}}{c^2}+\bs{v}\vecp\bs{B}\right]$ is the simultaneity--corrected Lorentz force per unit charge.

\subsection*{Case studies}
Consider a particle making a planar motion, as described in an inertial frame by the trajectory $\boldsymbol{r}_t$. 
Let $\boldsymbol{\Hat{v}}_t, \boldsymbol{\Hat{n}}_t, \boldsymbol{\Hat{b}}$ the Frenet--Serret apparatus for the curve $\boldsymbol{r}_t$. 
As shown in the Methods section, 
the simplest description of the motion of the spin is that in the comoving frame whose axes are Fermi--Walker (FW) transported \cite{Fermi,Walker}.  
The FW axes can be calculated, in the lab frame representation, for a general two--dimensional motion, as%
\bsubs
\bal
 {\vec{e}\,}'_{\!(1)}=&\begin{pmatrix}\gamma_t\cos{(\psi_t)} v_t\\  \gamma_t\cos(\psi_t)\boldsymbol{\Hat{v}}_t
 -\sin{(\psi_t)}\boldsymbol{\Hat{n}}_t\end{pmatrix},\\
{\vec{e}\,}'_{\!(2)}=&
\begin{pmatrix}\gamma_t\sin{(\psi_t)} v_t\\  \gamma_t\sin{(\psi_t)}\boldsymbol{\Hat{v}}_t+\cos{(\psi_t)}\boldsymbol{\Hat{n}}_t\end{pmatrix}, \\
{\vec{e}\,}'_{\!(3)} =&\begin{pmatrix}0\\ \boldsymbol{\Hat{b}}\end{pmatrix},
\eal
\label{eq:2dfw}%
\esubs
where $\psi_t$ is the angle $\psi_t=\int^t \gamma_t \kappa_t v_t dt$, with $\kappa_t$ the curvature of $\boldsymbol{r}_t$. 
The Thomas--Wigner rotation $\mathcal{R}_t$ can be calculated explicitly in this case, 
\beq
\mathcal{R}_t=
\bp
\cos(\psi_t-\alpha_t)&-\sin(\psi_t-\alpha_t)&0\\
\sin(\psi_t-\alpha_t)&\cos(\psi_t-\alpha_t)&0\\
0&0&1
\ep ,
\label{eq:twplanar}
\eeq
where $\alpha_t=\int^t  \kappa_t v_t dt$ is the Darboux rotation angle, \ie\ the angle that the tangent 
$\bs{\Hat{v}}_t$ makes with the $X$ axis. 

%
\subsubsection*{Uniform circular motion}
Let us consider the simple case of a particle, carrying an intrinsic angular momentum $\ushort{S}$, in uniform circular motion, so that $\gamma=const$ and $\kappa=const$. 
The Thomas--Wigner rotation $\mathcal{R}_t$ in this case is
\beq
\mathcal{R}_t=
\bp
\cos[(\gamma-1)\omega t]&-\sin[(\gamma-1)\omega t]&0\\
\sin[(\gamma-1)\omega t]&\cos[(\gamma-1)\omega t]&0\\
0&0&1
\ep, 
\label{eq:circrot}
\eeq
while the FW axes are 
\bsubs
\bal
 {\vec{e}\,}'_{\!(1)}=&\begin{pmatrix}\gamma\cos(\gamma \omega t)\, v\\  \gamma\cos(\gamma\omega t)\boldsymbol{\Hat{v}}_t
 -\sin{(\gamma\omega t)}\boldsymbol{\Hat{n}}_t\end{pmatrix},\\
{\vec{e}\,}'_{\!(2)}=&
\begin{pmatrix}\gamma\sin(\gamma\omega t)\, v\\  \gamma\sin{(\gamma\omega t)}\boldsymbol{\Hat{v}}_t+\cos{(\gamma \omega t)}\boldsymbol{\Hat{n}}_t\end{pmatrix}, \\
{\vec{e}\,}'_{\!(3)} =&\begin{pmatrix}0\\ \boldsymbol{\Hat{b}}\end{pmatrix}.
\eal
\label{eq:circfw}%
\esubs
Thus, for a motion with frequency $\omega$, a vector $\ushort{S}=X'_{t'} \ushort{e}_{(1)}+Y'_{t'} \ushort{e}_{(2)}+Z'_{t'} \ushort{e}_{(3)}$ with components 
$X'_{t'},Y'_{t'},Z'_{t'}$ with respect to the FW basis yields in the lab frame
a vector moving according to: 
\bsubs
\bal
\bs{S}=&\ \gamma\left[X'_{t'}\cos(\gamma \omega t)+Y'_{t'}\sin(\gamma \omega t)\right]\boldsymbol{\Hat{v}}_t
-\left[X'_{t'}\sin(\gamma \omega t)-Y'_{t'}\cos(\gamma \omega t)\right]\boldsymbol{\Hat{n}}_t
+ Z'_{t'} \bs{\Hat{b}} \ , \\
\bs{T}=&\ \gamma^{-1}\left[X'_{t'}\cos(\gamma \omega t)+Y'_{t'}\sin(\gamma \omega t)\right]\boldsymbol{\Hat{v}}_t
-\left[X'_{t'}\sin(\gamma \omega t)-Y'_{t'}\cos(\gamma \omega t)\right]\boldsymbol{\Hat{n}}_t
+ Z'_{t'} \bs{\Hat{b}}  ,
\eal
\label{eq:circspin}%
\esubs
where $t'=\tau(t)=t/\gamma$. 

We fixed the origin of time so that 
\beq
\bs{\Hat{v}}_t=\cos(\omega t) \bs{\Hat{\imath}}+\sin(\omega t) \bs{\Hat{\jmath}},\qquad  \bs{\Hat{n}}_t=-\sin(\omega t) \bs{\Hat{\imath}}
+\cos(\omega t) \bs{\Hat{\jmath}}.\eeq 

%
\subsubsection*{Spin with no magnetic moment}
We shall consider $\lambda=0$, so that the spin of the particle does not couple to the field causing the motion, and thus 
only the kinematic term contributes to its time evolution. 
Then, the terms $X', Y', Z'$ in Eqs.~\eqref{eq:circspin} are constant. 

\begin{figure}[h]
\begin{centering}
\includegraphics[width=0.99\textwidth]{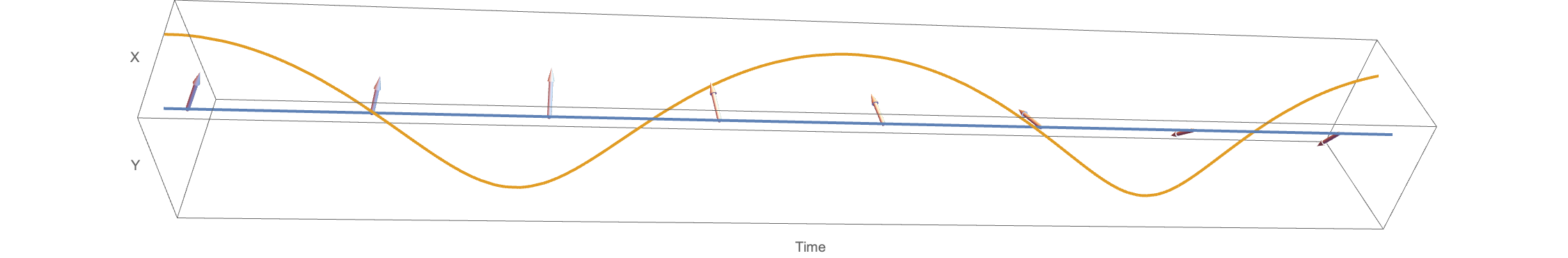}
\caption{\label{fig:spinex}
The parametric plot of the in-plane components $\bs{T}_\parallel(t)$ of the spin $\boldsymbol{T}$ carried along a point making a uniform circular motion, as seen from an inertial observer at the center of the motion. We used $v=3c/5$.}
\end{centering}
\end{figure}

A parametric plot of this solution is provided in Fig.~\ref{fig:spinex}. 
The squared norm $T^2$ of the simultaneity--corrected spin oscillates between the maximum, intrinsic value of the spin, $X'^2+Y'^2+Z'^2$ and the minimum $[X'^2+Y'^2]/\gamma^2+Z'^2$, obtained when the in--plane component is parallel to the velocity; 
correspondingly, the squared norm $S^2$ of the conventional spin reaches a minimum $X'^2+Y'^2+Z'^2$ and a maximum, $\gamma^2[X'^2+Y'^2]+Z'^2$, 
as illustrated in Fig.~\ref{fig:s2}. 

For the external electrons orbiting a nucleus the  velocity is approximately $v\simeq Z_\mathrm{eff}(n)\alpha c/n$, where $\alpha\simeq 1/137$ is the fine structure constant, $n$ the principal orbital number, and $Z_\mathrm{eff}(n)$ the effective nuclear number seen in the orbital $n$ because of the screening. The contraction effect is very small, of order $5{\vecp}10^{-5}$. For inner shell electrons of a heavy nucleus, where the effective charge of the nucleus is increased up to $Z$, the effect is a little larger, reaching 1 part in 1000.
{\begin{figure}[h]
\begin{centering}
\includegraphics[width=0.99\textwidth]{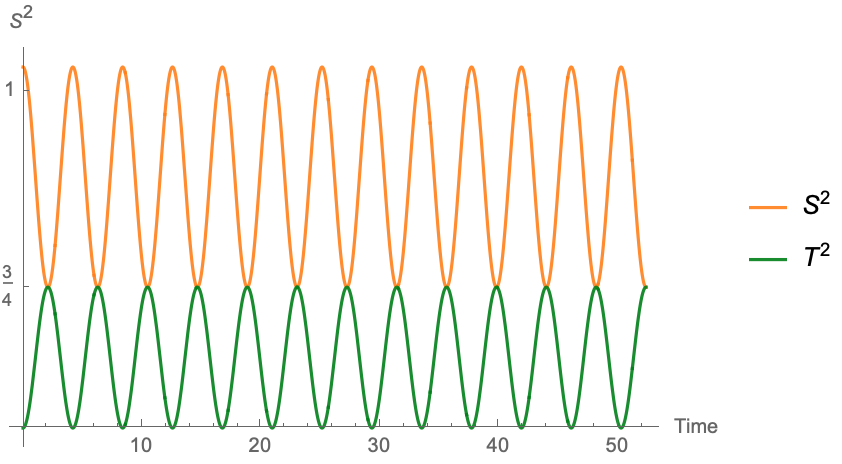}
\caption{\label{fig:s2} The variation in time of the total spin squared $\bs{S}^2$ and $\bs{T}^2$ divided by $\hbar^2$. For the sake of comparison, we considered the values $X'=1/\sqrt{2},Y'=0,Z'=1$, as in \cite[pp. 175--176]{MTW}. We put $v=3c/5$ to make the effect of Lorentz dilation/contraction evident. As discussed in the text, in an atom the speed of the orbiting electrons is too small compared to $c$ to allow to detect the effect.}
\end{centering}
\end{figure}}

The solution $\boldsymbol{S}$ is provided in \cite[pp. 175--176]{MTW}, as an exercise. 
The authors of \cite{MTW} arrive to the conclusion that the precession rate is $(\gamma-1)\omega=\gamma^2 v^2\omega/[c^2(\gamma+1)]$, by splitting the oscillation in a steady precession and a term that, it is argued, averages to zero. 

Indeed, according to the decomposition outlined in the Methods section, the linear part of the  angular velocity is 
$\boldsymbol{\omega}^\mathrm{Th}=\omega^\mathrm{Th}\boldsymbol{\Hat{b}}$, with 
$\omega^\mathrm{Th}=-(\gamma^2-1)\omega /2$, 
while the nonlinear part for the conventional ($+$ sign) and the simultaneity--corrected ($-$ sign) spin is $\bs{\omega}^{\mathrm{nl},\pm}=\omega^{\mathrm{nl},\pm} \bs{\Hat{b}}$
with 
\bal
{\omega}^{\mathrm{nl},\pm}=&\ 
\pm\frac{\gamma^2}{2} [\bs{a}_t\mycdot\bs{\Hat{P}}^\pm_t\, \bs{\Hat{P}}^\pm_t\vecp\bs{v}_t
+\bs{v}_t \mycdot \bs{\Hat{P}}^\pm_t\, \bs{\Hat{P}}^\pm_t\vecp\bs{a}_t]\mycdot\bs{\Hat{b}}
\nonumber \\
=&\ \pm \frac{\gamma^2-1}{2} 
\frac{(\gamma^{\pm 2}+1)\cos^2(\gamma\omega t-\phi)-1}{(\gamma^{\pm 2}-1)\cos^2(\gamma\omega t-\phi)+1}  \omega ,
\label{eq:omnl}
\eal
with $\tan(\phi)=Y'/X'$ and $\bs{\Hat{P}}^\pm$ the unit vector of the in--plane component of the spin.
Thus $\bs{\omega}^\mathrm{nl}$ oscillates with a fundamental frequency $2\gamma \omega$. 
The average value over a period is 
\beq
\overline{\omega}^{\mathrm{nl},\pm}=\frac{(\gamma-1)^2}{2}\omega \ ,
\eeq
thus it contributes a higher order correction, with respect to the linear term, in the non--relativistic limit. 
Remarkably, the average contribution is the same for both the conventional and the simultaneity--corrected spin. 

When this average correction is added to the linear part, we get, for a lucky coincidence, the accepted value of Thomas precession, 
\beq
\overline{\bs{\omega}}_\mathrm{tot}=-\frac{1}{2}\left[\gamma^2-1-(\gamma-1)^2\right] \omega \bs{\Hat{b}}=-(\gamma-1)\omega\bs{\Hat{b}} .
\eeq
Thus, we can say that the accepted result for Thomas precession holds, as an average, for uniform circular orbits, when adding the constant linear and the average nonlinear contribution to the precession. 
%
\subsubsection*{Spin with magnetic moment in a cyclotron \label{sub:subexample}}
In the lab frame, let $\bs{B}=B\bs{\Hat{b}}$ a uniform magnetic field and let the electric field $\bs{E}=\bs{0}$. 
Let the lab frame the frame in which the particle is making a uniform circular motion (and not a helix) with cyclotron frequency $\omega = -q B/\gamma m$ (a positive charge will rotate clockwise as seen from the positive semi-axis of the magnetic field, hence the minus sign). 
Contrary to the previous case, we do not put $\lambda=0$, but rather $\lambda=g_\mathrm{L} q/2m$, 
so that the term due to the magnetic field now contributes as well to the time--evolution of the spin. 
This case is relevant for muon and electron experiments \cite{muon,electroncyclotron,muon2}.

In the comoving FW frame the electromagnetic field is
\bal
\bs{B}'=\gamma \bs{B} , \qquad \bs{E}' = \mathcal{R}_{t'}(\gamma \bs{v}_t\vecp\bs{B}) ,
\label{eq:restemfield}
\eal
with $\mathcal{R}_{t'}$ given in Eq.~\eqref{eq:circrot}. 
Thus 
\beq
\bs{E}' = \gamma v B [
\sin(\gamma\omega t)\bs{\Hat{\imath}}-\cos(\gamma\omega t)\bs{\Hat{\jmath}}] .
\eeq
From the point of view of the particle, the force $q\bs{E}'$ due to the electric field is compensated by the force $m\bs{g}$ due to the local gravitational field \cite{Einstein1907} $\bs{g}=-\bs{a}'_\mathrm{prop}$, where $\bs{a}'_\mathrm{prop}$ is commonly called the proper acceleration (the particle however, does not feel an acceleration, since it is by definition at rest in its comoving frame: it feels a gravitational field, which has the dimension of an acceleration due to the equivalence of inertial and gravitational mass). 

In the comoving FW frame, according to Eq.~\eqref{eq:simple} derived in the Methods section, 
the spin precesses with angular velocity $g_\mathrm{L}\gamma^2 \omega/2$, 
\bsubs
\bal
X'_{t'}=&\ S'_\parallel \cos\left(\frac{g_\mathrm{L}}{2}\gamma^2\omega t'+\phi\right),\\
Y'_{t'}=&\ S'_\parallel \sin\left(\frac{g_\mathrm{L}}{2}\gamma^2\omega t'+\phi\right),\\
Z'_{t'}=&\ Z'_0 ,
\eal
\esubs
with $S'_\parallel = \sqrt{X'^2+Y'^2}=\sqrt{S'^2-Z'^2}$ the constant in--plane component of the spin, and $\phi$ a constant depending on the initial conditions.
Replacing these equations in Eq.~\eqref{eq:circspin}, with $t=\gamma t'$, we obtain the time evolution in the lab frame for a magnetic moment in a cyclotron, 
\bsubs
\bal
\bs{S}=&\  S'_\parallel \left[\gamma\cos(a \gamma \omega t+\phi)\boldsymbol{\Hat{v}}_t
+\sin(a\gamma \omega t+\phi)\boldsymbol{\Hat{n}}_t\right]
+ Z' \bs{\Hat{b}} \ , \\
\bs{T}=&\  S'_\parallel \left[\gamma^{-1}
\cos(a \gamma \omega t+\phi)\boldsymbol{\Hat{v}}_t
+\sin(a\gamma \omega t+\phi)\boldsymbol{\Hat{n}}_t\right]
+ Z' \bs{\Hat{b}} \ ,
\eal
\label{eq:circspin2}%
\esubs
with $a=(g_\mathrm{L}-2)/2$ the $g$--factor anomaly. 
Thus, if the $g$--factor were exactly 2, \ie\ $a=0$, the spin would stay rigidly fixed with respect to the Frenet--Serret apparatus formed by the tangent,  the normal, and the binormal of the trajectory. 

The linear precession rate is 
\beq
\bs{\omega}^\mathrm{Th} = [1+\tfrac{1}{2} (\gamma^2+1)a] \omega \bs{\Hat{b}}, 
\eeq
while the nonlinear precession rate is 
\beq
\bs{\omega}^{\mathrm{nl},\pm} = \mp\frac{1}{2}(\gamma^2-1) a\omega \frac{(\bs{\Hat{v}}_t\mycdot\bs{\Hat{P}}^\pm_t)^2-(\bs{\Hat{n}}_t\mycdot\bs{\Hat{P}}^\pm_t)^2}
{(\bs{\Hat{v}}_t\mycdot\bs{\Hat{P}}^\pm_t)^2+(\bs{\Hat{n}}_t\mycdot\bs{\Hat{P}}^\pm_t)^2}  \bs{\Hat{b}}, 
\label{eq:nlcirc}%
\eeq
where as in the previous case $\bs{P}^\pm$ stands for the conventional or the simultaneity corrected spin, depending on the index being $+$ or $-$, respectively. 

After replacing Eqs.~\eqref{eq:circspin2} into Eq.~\eqref{eq:nlcirc}, we have 
\beq
\bs{\omega}^{\mathrm{nl},\pm} = \mp\frac{1}{2}(\gamma^2-1)a \omega 
\frac{(\gamma^{\pm 2}+1)\cos^2(a\gamma\omega t+\phi)-1}{(\gamma^{\pm 2}-1)\cos^2(a\gamma\omega t+\phi)+1}   \bs{\Hat{b}} .
\eeq
A time-average over a period $T=\pi/a\gamma\omega$ yields 
\beq
\overline{\bs{\omega}}^{\mathrm{nl}\pm} =-\frac{1}{2}a(\gamma-1)^2\omega \bs{\Hat{b}} ,
\eeq
so that the overall average precession is 
\beq
\overline{\bs{\omega}}_\mathrm{tot}
= (1+a\gamma)\omega \bs{\Hat{b}}.
\label{eq:muon}
\eeq
As in the former case, this is the value obtained using the incorrect formula for Thomas precession. 

The fluctuations are 
\beq
\Delta \omega = \sqrt{\frac{\gamma}{2}}(\gamma-1)a\omega .
\eeq

\subsubsection*{Non-circular orbits}
For a general orbit where the linear precession rate  
$(\gamma_t^2/2) \bs{v}_t\vecp\bs{a}_t$ is not constant, adding 
an average contribution $\overline{\bs{\omega}}^\mathrm{nl}$ can not lead to the accepted formula $\bs{\omega}_\mathrm{tot}=-[\gamma^2_t/(\gamma_t+1)]\bs{v}_t\times \bs{a}_t$. 
As an example, we take the relativistic orbits in a Coulomb potential, as discussed by Darwin and Sommerfeld
\cite{Darwin1913,Sommerfeld}. 

For simplicity, we will consider $\lambda=0$. 
After a few calculations, the Thomas precession term is found to be
\beq
\omega^\mathrm{Th}=-\frac{h}{2\rho^3}=-\frac{L e^2}{2m^2 c^2r^3} ,
\eeq
where 
$\rho$ the distance from the center $r$ in units 
of the electron radius, $R_0=e^2/mc^2$. 
while $h$ is the angular momentum $L$ in units of $mR_0c$, with 
$e$ the charge of the electron, and $m$ its mass.

The non--linear precession has a more involved expression, which we will provide for the simultaneity--corrected 
spin only:
\beq
\omega^\mathrm{nl,-}=-\frac{u^2}{2}\frac{\frac{hu}{(\epsilon+u)^2}\cos^2{\psi_u}-hu\sin^2{\psi_u}
+ 2 \frac{\sqrt{Q^*_u}}{(\epsilon+u)^3} \sin{\psi_u}\cos{\psi_u}}{\frac{1}{(\epsilon+u)^2}\cos^2{\psi_u}+\sin^2{\psi_u}},
\eeq
where $u=1/\rho$, $\epsilon$ is the total energy in units of the rest energy $mc^2$, while $Q^*_u=\epsilon^2-1+2\epsilon u+(1-h^2)u^2$ is a quadratic form in $u$, and 
\bal
\psi_u &=  \frac{h}{h^2-1}\sqrt{Q^*_u}+\frac{h\epsilon}{(h^2-1)^{3/2}}
\arctan\left[\frac{\epsilon+(1-h^2)u}{\sqrt{(h^ 2-1)Q^*_u}}\right]
\nonumber 
\\
&-\frac{1}{2}
\arctan\left[\frac{(\epsilon-1)(1+h^2u)+u}{(\epsilon-1)h\sqrt{Q^*_u}}\right]
-\frac{1}{2}\arctan\left[\frac{(\epsilon+1)(1+h^2u)+u}{(\epsilon+1)h\sqrt{Q^*_u}}\right] .
\eal
%
\section*{Discussion}
In conclusion, the accepted value for the Thomas precession stems from confusing it with the rate of Thomas--Wigner rotation, which applies in a rotating frame possibly convenient but of no physical significance.  

Instead, we wish to know the equations of motion of the spin either in the comoving frame the axes of which undergo FW transport, or 
in the lab frame. In the former frame, the equation takes its simplest form, Eq.~\eqref{eq:simple}, in the latter frame either 
Eq.~\eqref{eq:laba} or \eqref{eq:labb} holds, depending if one is using the conventional definition of spin $\bs{S}$ 
or the simultaneity--corrected one $\bs{T}$. 

The reason behind the complicated motion in the lab frame is that a Lorentz boost introduces a rotation: indeed, if a vector $\bs{S}'$ 
stays constant in the rest frame of the particle, when going to the lab frame one has to apply a Thomas--Wigner rotation and a boost. 
The boost changes the component of $\bs{S}$ along the velocity, thus 
the vector will undergo an additional rotation. This rotation is nonlinear, as it depends on the orientation of $\bs{S}'$ relative to the velocity. In Section C of the Supplemental Materials we show in detail how the contribution of the Thomas--Wigner rotation and the contribution of Lorentz contraction get twined. 

In a uniform circular motion, the effect of the Lorentz contraction averages out, because the Lorentz contraction factor is constant and because the spin, over time, points in all directions. This exceptional case, however, has been mistaken for the norm, since the uniform circular motion is the simplest curvilinear accelerated motion, and hence is the go--to case when treating Thomas precession. 

While in the lab frame the spin does not undergo a simple precession, we have shown how to extract univocally from the equation of motion a linear rotation term, which we identify with the Thomas precession, and a nonlinear rotation term, due to the interplay of Lorentz contraction and Thomas--Wigner rotation. 

In perspective, it is interesting to consider the expansion of the present results to the general relativistic case, along the lines of Ref.~\cite{Chakraborty2017}, where the Thomas precession combines with de Sitter and Lense--Thirring precession. 

%

\section*{Methods}
%
\subsection*{Relativity of simultaneity}
Given the 4--vector field $\ushort{S}_t$ along the trajectory of a particle $O'$, what is the spin, as a three--dimensional vector, that an observer $O$ will assign? 
The conventional procedure is to consider the representation, in the frame of $O$, $\vec{S}=\{S^0,\bs{S}\}$, and take the spatial part 
$\bs{S}$ as the spin. (We have introduced the symbol $\vec{S}$ for the contravariant representation, in a given reference frame, of the abstract vector 
$\ushort{S}$. 
We shall keep the concepts of a vector and of its representation distinct, as explained in the Supplemental Materials.)
However, since $O$ and $O'$ are in relative motion, the relativity of simultaneity applies, which results in a different 3--vector $\bs{T}$ to be associated 
by the observer $O$.  

{\begin{figure}[h!]
\begin{centering}
\includegraphics[width=0.9\textwidth]{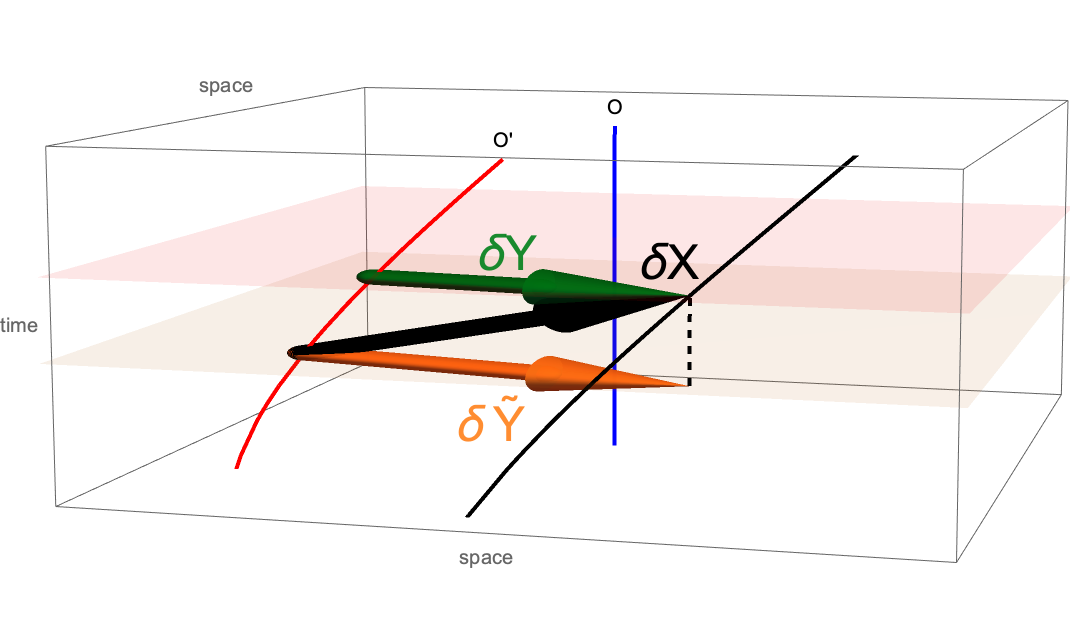}
\caption{\label{fig:two} There are two procedures that an inertial observer can use to build a vector field normal to the worldline $O$ from the vector field normal to $O'$ (black arrow): by projecting it on the simultaneity eigenplane passing through the origin of the vector (orange), or by building a vector joining the points at which the hyperplane of simultaneity crosses $O'$ and the line traced by the tip. We note that the time component gives a negative contribution to the squared length of a vector, thus the orange vector is larger than the black vector (by a Lorentz factor $\gamma$, to be precise). On the other hand, the shortening in space of the green vector more than compensates for its increase due to the suppression of the time component, yielding a contraction by a factor $\gamma$ instead.}
\end{centering}
\end{figure}}

To illustrate this point, we consider an infinitesimal vector in Minkowski space, $\ushort{\delta X}_{t'}$, transported along $O'$. 
The tip of this vector points to $P'_{t'}=O'_{t'}+\ushort{\delta X}_{t'}$ from the point $O'_{t'}$. 
The tip of the vector thus traces a curve $P'$ close to $O'$. 
For an inertial observer $O$, whose worldline is a straight line, the spatial vector to associate is 
$\ushort{\delta Y}_t$, obtained by intersecting the two curves $O'$ and $P'=O'+\ushort{\delta X}$ with a hyperplane of simultaneity $t=const$, 
then considering the infinitesimal vector from the intersection $O'_t$ to the intersection $P'_t$. 
By construction, in the inertial frame the representation of $\ushort{\delta Y}_t$ is purely spatial $\vec{\delta Y}=\{0,\bs{\delta Y}_t\}$. 

If $\ushort{\delta X}$ is a rod of intrinsic 
length $\delta l=\sqrt{\ushort{\delta X}\mycdot\ushort{\delta X}}$ carried by the accelerated observer, the corresponding vector 
$\bs{\delta Y}$ describes a rod contracted by the Lorentz factor $|\bs{\delta Y}|=\delta l/\gamma$. 
If, instead, we applied the conventional procedure to a rod, rather than contracting, it would dilate in the direction of motion, $|\bs{\delta X}|=\gamma\delta l$,
as illustrated in Fig.~\ref{fig:two}. 

The procedure accounting for the relativity of simultaneity applies to any vector field transported normally to a worldline $O'$. 
We shall refer to it as the simultaneity-corrected procedure. 
Two other important vector fields normal to the worldline of a particle are its 4--acceleration and the 
Lorentz 4--force on it. In a frame where their representation is $\{A^0,\bs{A}\}$ and $\{F^0,\bs{F}\}$, 
the 3--vectors conventionally associated with these fields are 
$\bs{A}=\gamma^2 \bs{a}+\frac{\gamma^4}{c^2}\bs{a}\mycdot\bs{v}\,\bs{v}$ and 
$\bs{f}=\bs{F}/\gamma=\bs{E}+\bs{v}\vecp\bs{B}$. 
We note how the definitions are mutually inconsistent, due to the extra $1/\gamma$ factor in the force $\bs{f}$, which is introduced in order to keep the form of Newton's second law $\bs{f}=\frac{d}{dt} \bs{p}$. 

The simultaneity corrected 3--vectors are 
instead $\bs{A}^\mathrm{sim}=\gamma^2\bs{a}$ and 
$\bs{G}=\gamma\left[\bs{E}-\bs{E}\mycdot\bs{v}\,\frac{\bs{v}}{c^2}+\bs{v}\vecp\bs{B}\right]$, with 
$\bs{E},\bs{B}$ the electric and magnetic field. Newton's second law becomes then: $q\bs{G}=m\bs{A}^\mathrm{sim}$. 

In the case of spin, we opt for the simultaneity--corrected procedure because the intrinsic magnetic moment of the electron, 
in the Gilbert model, might be due to two equal and opposite 
magnetic monopoles separated by a small distance. (Precisely, if we take into account Dirac's quantization and take the monopoles to be elementary, the separation explaining the observed magnetic dipole moment is the classical radius of the electron $\delta r=e^2/(mc^2)$.) Thus, the magnetic moment should behave as a rod. 

If instead the magnetic moment is due to the rotation of a charge, as in the Amp\`{e}re model, consider the following: 
a distributed charge is making a circular motion around its center of mass in a given frame. 
In a frame where the center of mass moves with speed $\bs{v}$ normal to the plane of circular motion, the charges will make 
a helical trajectory, while their angular speed is reduced by a factor $\gamma$ due to the dilation of time, and the magnetic moment is reduced correspondingly. 

The magnetic moment, therefore, decreases by a factor $1/\gamma$ in the direction of motion, irrespective of the underlying model.

A further argument in favour of using the simultaneity--corrected spin $\bs{T}$ is the following. The magnetic dipole moment $\bs{M}$ 
is obtained by the integration over space of the magnetic dipole density $\bs{m}$. The latter is a 3--vector, which, together with the electric dipole density $\bs{p}$, forms the skew--symmetric electromagnetic dipole tensor $d=(\bs{p},\bs{m}):=\bp0&\bs{p}^\tr\\\bs{p}&\bs{m}\cdot\bs{\mathcal{J}}\ep$. (We are representing $d$ as a matrix, $d^\alpha_{\,\beta}$, \ie\ a contravariant--covariant tensor, thus the skew--symmetry is not reflected in the components mixing time and space). 
The electric and magnetic dipole density 4--vectors are obtained by $d$ and its dual 
$*d=\bp0&\bs{m}^\tr\\\bs{m}&-\bs{p}\cdot\bs{\mathcal{J}}\ep$ by contraction with the 4--velocity, $p^\alpha=d^\alpha_{\,\beta}U^\beta$, $m^\alpha=*d^\alpha_{\,\beta}U^\beta$, 
hence they are automatically normal to the 4--velocity. 
In the rest frame of the electron $\bs{p}'=0$, as no intrinsic electric dipole of the electron has been observed so far. 
Therefore, in the lab frame, recalling that the Lorentz transformation from the lab frame to the electron rest frame includes the Thomas--Wigner rotation, $L_t=R_t B[\bs{v}_t]$, with $B[\bs{v}]$ a Lorentz boost, the magnetic dipole moment density is 
$\bs{m}= \mathcal{R}_t^\tr\bs{m}'-(\gamma_t-1)\bs{\Hat{v}}_t\vecp(\bs{\Hat{v}}_t\vecp \mathcal{R}_t^\tr\bs{m}')$. The component of $\mathcal{R}_t^\tr\bs{m}'$ parallel to the velocity is unaffected by the Lorentz contraction, $\bs{\Hat{v}}_t\mycdot\bs{m}= 
\bs{\Hat{v}}_t\mycdot(\mathcal{R}_t^\tr\bs{m}')$, while the component perpendicular to the velocity is multiplied by a factor $\gamma$. Upon integration over the volume occupied by the electron, we get finally 
\beq
\bs{M} = \int dV \bs{m}= \int \frac{dV'}{\gamma_t}[\mathcal{R}_t^\tr\bs{m}'+(\gamma_t-1)\bs{\Hat{v}}_t\vecp(\bs{\Hat{v}}_t\vecp \mathcal{R}_t^\tr\bs{m}')]=\frac{\mathcal{R}_t^\tr \bs{M}'}{\gamma_t}+\frac{\gamma_t-1}{\gamma_t}\bs{\Hat{v}}_t\vecp(\bs{\Hat{v}}_t\vecp \mathcal{R}_t^\tr\bs{M}').
\eeq
Thus, the total dipole magnetic moment contracts by a factor $\gamma$ in the direction of the velocity. 
In the rest frame of the electron $\bs{M}'=\lambda \bs{S}'$, with $\lambda = g_\mathrm{L} e/(2m)$, $e=-|e|$ being the signed charge of the electron. 
In order for the same proportionality to apply in the lab frame, 
we must have $\bs{M}=\lambda\bs{T}$, \ie\ the simultaneity--corrected spin must be used.

As a test of which definition is relevant, consider a beam 
of electrons at relativistic speed. 
It is possible to polarize the electrons longitudinally, so that their spin is parallel to their velocity and their magnetic moment is antiparallel to it. 
If the beam enters a region of magnetic field $B$ parallel to the spin, a Zeeman splitting occurs, with the electrons in the upper level, along with a reduction of the initial speed due to the magnetic field gradient. 
If the spin, and hence the magnetic moment, were increased by a factor $\gamma$, then the deceleration of the electrons should be accordingly large, and also the frequency of the photons emitted in the relaxation to the lower Zeeman state should be large. 
For instance, in a 1GeV linear accelerator, electrons can reach a factor $\gamma\simeq 2\times 10^3$, making the effect, or the lack thereof, easily detectable. 

We denoted the 3--dimensional vector field constructed according to the relativity of simultaneity as $\bs{T}$, 
\beq
\bs{T} = \bs{S}-S^0 \bs{v} =  \bs{S}- \bs{S}\mycdot\bs{v}\, \bs{v} \ .
\eeq
From here on, we shall use units where $c=1$. 
We also note that, in an inertial frame, the scalar product of the two definitions yields the constant squared norm of the spin 
\beq
\bs{S}\mycdot\bs{T} = \ushort{S}\mycdot\ushort{S} = \frac{3}{4}\hbar^2 .
\eeq


\subsection*{Fermi--Walker transport\label{sec:fwtransport}}
Fermi--Walker (FW) transport \cite{Fermi,Walker} has a crucial role in the following. 
We illustrate it right away.

Consider a particle, having a velocity $\bs{v}_{t_0}$, relative to an inertial frame 
$\mathfrak{S}$, at time $t_0$ in the inertial frame. Let us pick an inertial reference frame $\mathfrak{S}'_{t_0}$ where the particle is at rest at time $t_0$, and let $t'_0$ the proper time. 
At an infinitesimally later time $\delta t'$ in 
$\mathfrak{S}'_{t_0}$, the velocity of the particle changes by $\delta \bs{v}' = \bs{a}'_{t'_0} \delta t'$, with $\bs{a}'_{t'_0}$ 
the proper acceleration.  
Now, there are infinitely many different inertial reference frames where the particle is at rest at time $t'_0+\delta t'$, which can be obtained by applying an infinitesimal Lorentz boost followed by an arbitrary rotation of the spatial axes. A Lorentz boost does not rotate the spatial axes normal to the velocity, and it rotates the time axis and the spatial axis parallel to the velocity in their common plane. 
It is therefore the simplest Lorentz transformation. 
(Lorentz boosts are the transformations often presented in introductory texts on special relativity as the  Lorentz transformations \emph{tout court}. Actually, they form a subset, which is not a subgroup, 
 of Lorentz transformations: in special relativity, the 3D space is a hyperplane normal to the worldline of 
 the origin of a reference frame. Two reference frames in relative motion will thus have different 3D spaces. 
 These spaces, however, will share a common plane. Lorentz boosts are transformations between reference frames in which two of the spatial axes, usually the $Y$ and $Z$ axes, belong to the common plane and coincide between the two frames.) 

Thus, let us consider, among these infinitely many frames, the one which is obtained by a pure infinitesimal boost with velocity $\bs{\delta v}'$. Let us repeat the procedure at another infinitesimal time $\delta t''$, \etc. The sequence of infinitesimal boosts defines FW transport. Since the boosts do not form a subgroup, their product is a general Lorentz transformation, which can be written
as the product of a 3D rotation and a boost. 

In particular, if the spatial axes of the reference frame comoving with the particle are FW transported, they individuate a special accelerated reference frame, the FW frame, which is non--rotating. 
This means that, if we spin three gyroscopes attached to the particle, then accelerate the particle without applying any torque to the gyroscopes, the axes of these will be FW transported. While the gyroscopes do not rotate in the sense that a simple Lorentz boost is applied from an instantaneous comoving frame to the next, when viewing them from the initial reference frame $\mathfrak{S}$, they will make a complicated motion in which we shall individuate a linear precession term, the Thomas precession. 
It is erroneous, however, to state that the gyroscope will just appear as precessing in the lab frame. 
On one hand, Lorentz contraction will affect the gyroscopes, on the other hand, due to the relativity of simultaneity, the angles among them will appear to change in time in the lab frame, as shown in Fig.~\ref{fig:fwvsboost}. 

{\begin{figure}[h!]
\begin{centering}
\begin{subfigure}[t]{0.33\textwidth}
\includegraphics[width=\textwidth]{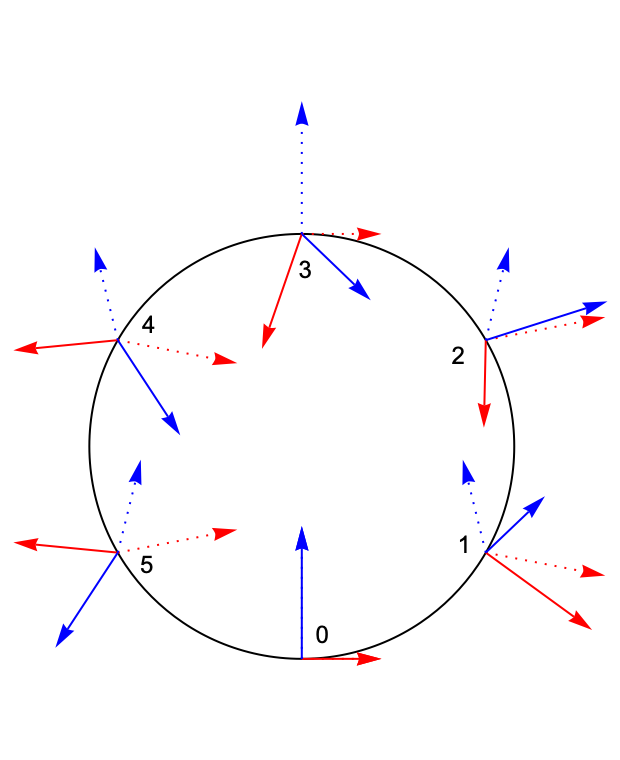}
\caption{}
\end{subfigure}
\begin{subfigure}[t]{0.3\textwidth}
\includegraphics[width=\textwidth]{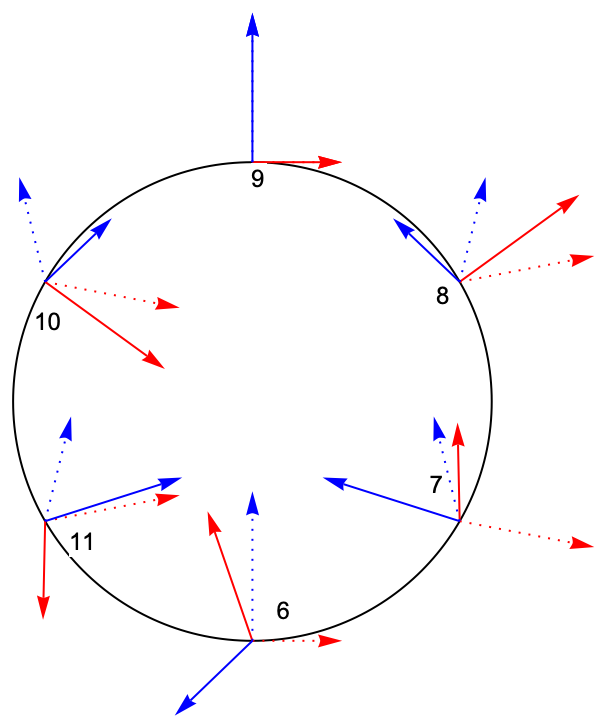}
\caption{}
\end{subfigure}
\begin{subfigure}[t]{0.3\textwidth}
\includegraphics[height=\textwidth, width=\textwidth]{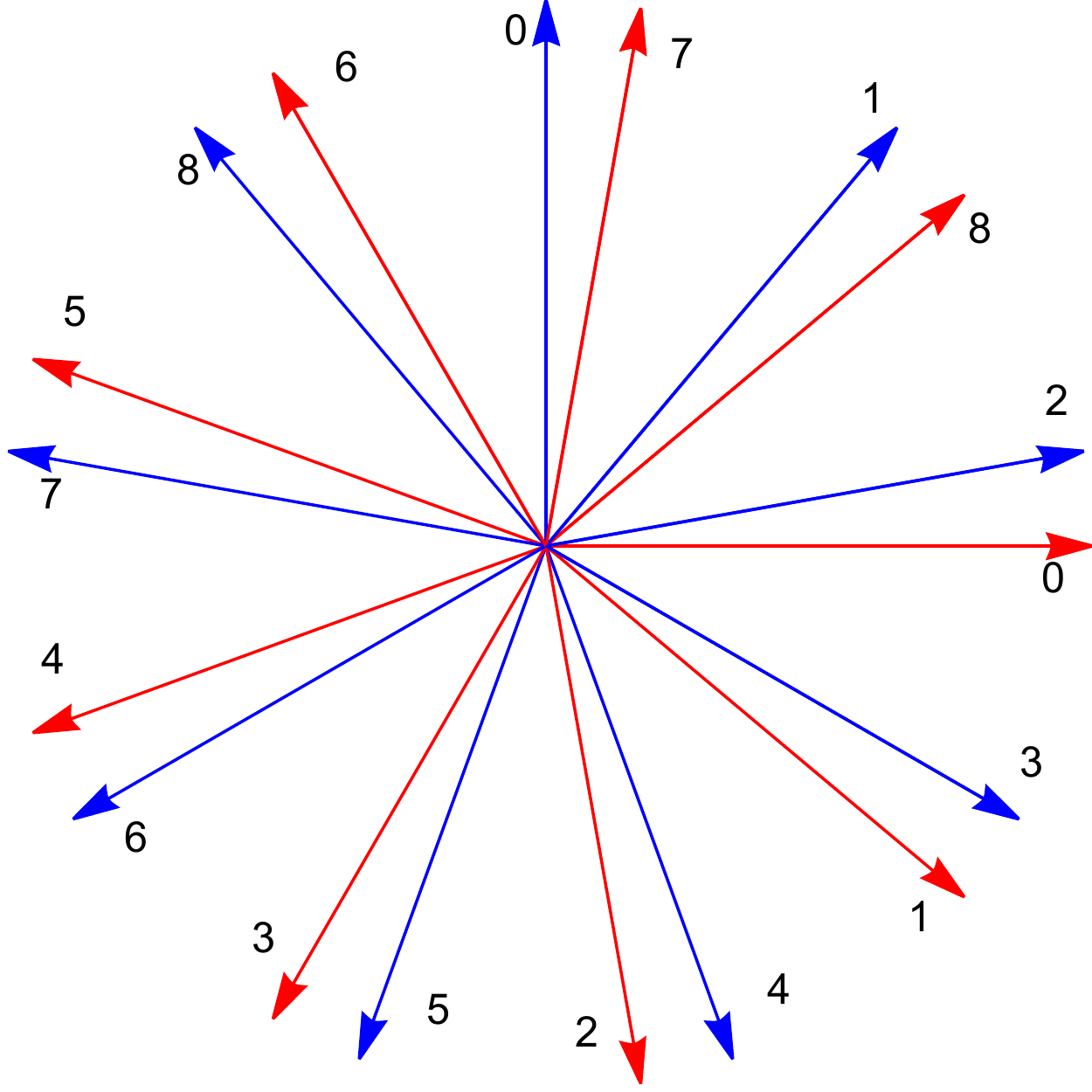}
\caption{}
\end{subfigure}
\caption{\label{fig:fwvsboost} (a) and (b). 
For a particle in uniform circular motion, the $X$ (red) and $Y$ (blue) axes 
will make a complicated motion as seen from the lab frame. In the article, we show how to extract a linear rotation, which we identify with Thomas precession, and an additional nonlinear rotation. The solid lines represent the axes of a Fermi--Walker accelerated frame, the dotted lines the axes of a boosted frame, 
defined by pure Lorentz boosts applied to the lab frame. 
The length of the axes is constant in the accelerated frame, but varies in the lab frame due to Lorentz 
contraction. The speed is $v=4c/5$, thus $\gamma=5/3$.
The axes were chosen to coincide with the lab axes at $t=0$, when 
the particle is at $\{0,-R\}$. The position of the axes is shown at times $t=n T/6$, with $T$ the period 
in the lab frame, for (a) $n=0,1,\dots,5$ and (b) $n=6,7,\dots, 11$. 
(c) In the frame defined by the dotted axes, the boosted frame, the solid axes 
appear to be purely rotating, with a frequency $\omega'=\gamma(\gamma-1)\omega$, hence a period $T'=9T/10$. 
Due to the time dilation, in the boosted frame the axes of the 
FW frame will appear as shown in the subfigure at times $t'=nT/10=nT'/9$.}
\end{centering}
\end{figure}}

In any other accelerated frame following the same worldline, but with axes not being FW transported, an additional Coriolis rotation term appears, causing an apparent precession of the gyroscopes in the accelerated frame in absence of external torques. In this sense, the FW frame is minimally non--inertial. 
The Fermi--Walker (FW) reference frame, $\mathfrak{S}'_{t}$, thus has a special status among the frames comoving with an accelerated particle: no Coriolis forces appear in it. 
More details can be found in the Supplemental Materials.  

The Lorentz transformation from the lab frame to the FW frame $L_\mathrm{FW}$ can be decomposed either as 
$R_\mathrm{TW}(t) B[\bs{v}_t]$ or as $B[\mathcal{R}_\mathrm{TW}(t)\bs{v}_t] R_\mathrm{TW}(t)$. 
The $4\times 4$ matrix $R_\mathrm{TW}(t)$ is a rotation in 3D space, the Thomas--Wigner rotation, its nontrivial part being the $3\times 3$ orthogonal matrix $\mathcal{R}_\mathrm{TW}(t)$. The $4\times 4$ matrix $B[\bs{v}_t]$ is a Lorentz boost. 

While the matrix $R_\mathrm{TW}(t)$ is the same in both decompositions, it represents two distinct physical rotations. 

In the decomposition $L_\mathrm{FW}= R_\mathrm{TW}(t) B[\bs{v}_t]$, the rotation is applied 
after the Lorentz boost, hence it represents the rotation of the axes of the FW frame relative to another frame, comoving with the particle, defined by the application of Lorentz boosts from the lab frame: the boosted rest frame, $\mathfrak{S}'^*_{t}$. 
The rate of rotation $\Omega'$ is obtained by differentiating in the proper time $t'$, 
$\frac{d}{dt'}\mathcal{R}_\mathrm{TW}(\tau^{-1}(t'))= \Omega'(t')\mathcal{R}_\mathrm{TW}(\tau^{-1}(t'))$. 

In the decomposition $L_\mathrm{FW}= B[\mathcal{R}_\mathrm{TW}(t)\bs{v}_t] R_\mathrm{TW}(t)$, the rotation is applied 
before the Lorentz boost, hence it represents the rotation of the axes of a frame whose origin is at rest in the lab frame, and whose axes are rotating relative to the lab frame. 
The rotation rate, as measured in the lab frame, is the skew--symmetric matrix $\Omega$ defined by 
$\frac{d}{dt}\mathcal{R}_\mathrm{TW}(t)= \Omega(t)\mathcal{R}_\mathrm{TW}(t)$. 
The two rotation rates satisfy $\Omega(t)=\Omega'(\tau(t))/\gamma(t)$. 

The relation between the proper time $t'$ of the accelerated particle $O'$ and the time $t$ of the lab is given, up to arbitrary choices of initial time, 
by the well known time dilation formula $d t'/d t = \gamma^{-1}_t$, where $\gamma_t=[1-v^2_t]^{-1/2}$ is the Lorentz contraction factor. 
For a given motion of $O'$, one finds, after integrating, $t'=\tau(t)$, and upon inversion $t=\tau^{-1}(t')$. 
Relativistic invariant formulas are independent of the reference frame, hence they are expressed in terms of proper time. 
We will use the time dilation formula and pass to the inertial time $t$ when representing the formulas in the lab frame, as we did in the preceding paragraph. 
Sometimes, we will write equations depending on $t'$ on one side and on $t$ on the other side. In these cases, for brevity, 
we will sometimes omit to write $t=\tau^{-1}(t')$ or $t'=\tau(t)$.

The 3--vector corresponding to the skew--symmetric matrix $\Omega'_{t'}=\bs{\omega}'^*_{t'}\mycdot\bs{\mathcal{J}}$, with $\bs{\mathcal{J}}$ the three $(3\times3)$ generators of rotations in 3D, is 
\beq
\bs{\omega}'^*_{t'}=
\frac{\gamma_{t'}}{\gamma_{t'}+1} \bs{v}'^*_{t'}\vecp\bs{a}'^*_{t'}=-\frac{\gamma^3_{t}}{\gamma_{t}+1} \bs{v}_{t}\vecp\bs{a}_{t},
\label{eq:spec1b}
\eeq
where $\bs{v}'^*_{t'}$ is the velocity of the lab frame relative to the boosted frame, $\gamma_{t'}$ is the corresponding Lorentz factor, and 
$\bs{a}'^*_{t'}$ is the proper acceleration of the boosted frame. 
The second equality specifies the rotation in terms of quantities measured in the lab frame. 

The vector associated to the skew--symmetric matrix $\Omega_t$ is 
\beq
\bs{\omega}_{t}=
-\frac{\gamma^2_{t}}{\gamma_{t}+1} \bs{v}_{t}\vecp\bs{a}_{t}.
\label{eq:spec1d}
\eeq

We recognize in Eq.~\eqref{eq:spec1d} the accepted value for the Thomas precession: it is not, therefore, an additional rotation of the spin in the lab frame, but rather the rotation of an auxiliary purely rotating reference frame. 


%

%
\subsection*{Equation of motion for the spin}
\subsubsection*{Invariant equation}
The fundamental invariant equation for $\ushort{S}$ must be determined. 
In the following, a double underline indicates a rank--2 tensor, \eg\ $\dunderline{F}$. A rank--2 tensor
 accepts two vectorial arguments, one to the left and one to the right. 
By juxtaposing two vectors $\ushort{A}\ushort{B}$ we mean a rank--2 tensor which can be multiplied to the left or to the right, \eg\ if 
$\dunderline{C}=\ushort{A}\ushort{B}$, 
$\ushort{V}\mycdot\dunderline{C}=\ushort{V} \mycdot \ushort{A} \ushort{B}$, 
$\dunderline{C}\mycdot\ushort{V}=\ushort{A}  \ushort{B}\mycdot\ushort{V}$.

Assuming a linear first order equation for a 4--vector $\ushort{S}$, 
its motion is necessarily of the form ($\partial_{t'} = d/d{t'}$)
\beq
\partial_{t'}\ushort{S}_{t'} = \lambda\dunderline{\Phi}_{t'}\mycdot\ushort{S}_{t'}+\dunderline{\mathfrak{B}}_{t'}\mycdot\ushort{S}_{t'} ,
\label{eq:genevol}
\eeq
with $\lambda$ a coupling constant, $\dunderline{\Phi}_{t'}$ an applied tensor field and $\dunderline{\mathfrak{B}}_{t'}:=
\ushort{U}_{t'} \ushort{A}_{t'}-\ushort{A}_{t'} \ushort{U}_{t'}$ the generator of FW transport,
$\ushort{U}_{t'}$ being the 4--velocity of $O'$ and 
$\ushort{A}_{t'}:=\partial_{t'} \ushort{U}_{t'}$ being its 4--acceleration.  

The symbol $\dunderline{\mathfrak{B}}$ for 
$\ushort{U}\wedge\ushort{A}$ was chosen to stress that the FW generator is the Minkowski space analog 
of the unnormalized binormal vector of the Frenet--Serret apparatus in 3--dimensional space, $\kappa\bs{\Hat{b}} = \bs{\Hat{v}}\vecp\bs{\Hat{n}}$, 
$\kappa$ being the curvature and $\bs{\Hat{n}}=\frac{d\bs{\Hat{v}}}{dl}$ the first normal. In Minkowski space, the unit tangent vector $\bs{\Hat{v}}$ corresponds to the 4--velocity $\ushort{U}$, the curvature to the norm of the 4--acceleration $\ushort{A}$, the first normal 
$\bs{\Hat{n}}$ to the direction of the 4--acceleration $\ushort{A}$, and the
curve length parameter $l$ to the proper time.

For a spin $1/2$ interacting with the electromagnetic field, $\lambda$ 
is the gyromagnetic factor, \ie\ the constant of proportionality between the spin and the intrinsic magnetic moment, which is the Bohr magneton times the Land\'{e} $g_\mathrm{L}$ factor divided by the total angular momentum, 
$\lambda = g_\mathrm{L} q/(2m)$.    
The skew--symmetry of $\dunderline{\mathfrak{B}}$ ensures the conservation of the norm of $\ushort{S}$ in absence of external fields. 
The first term on the rhs of Eq.~\eqref{eq:genevol} would be zero in the absence of fields acting directly on 
$\ushort{S}$. Hence, we shall refer to the terms arising from it as the direct part. 


If, furthermore, the field is 
$\dunderline{\Phi}_{t'}= \dunderline{\Pi}_{t'}\mycdot\dunderline{F}_{t'}\mycdot\dunderline{\Pi}_{t'}$, with 
$\dunderline{F}_{t'}$ skew--symmetric and $\dunderline{\Pi}_{t'}=1+\ushort{U}_{t'}\ushort{U}_{t'}$, the projection operator on the three--dimensional space normal to the 4--velocity $\ushort{U}_{t'}$, then 
Eq.~\eqref{eq:genevol} preserves both the norm of $\ushort{S}$ and the value of its projection along the worldline $O'$, 
$\ushort{S}_{t'}\mycdot\ushort{S}_{t'}=const$ and $\ushort{U}_{t'}\mycdot\ushort{S}_{t'}=const$. 
In particular, when the constant projection is zero, the field $\dunderline{F}$ is the electromagnetic field, 
and the acceleration is due solely to $\dunderline{F}$ (neglecting the Stern--Gerlach contribution to the acceleration \cite{reldynmagnet})  
$m\ushort{A}_{t'}=q\dunderline{F}_{t'}\mycdot \ushort{U}_{t'}$, then Eq.~\eqref{eq:genevol} reduces to the Bargmann--Michel--Telegdi equation \cite{BMT}. 
\subsubsection*{Equation for components in an arbitrary frame}
A reference frame is characterized by an origin $O$ following a given worldline in Minkowski space, and by a family of three vectors 
$\ushort{e}_{(j)}(t)$ normal to the worldline, the spatial axes. 
(To avoid confusion, we are writing the time dependence in conventional form, 
$\ushort{e}_{(j)}(t), S^j(t)$ when indexes appear. 
In the following, however, we will avoid as far as possible the use of component indexes. All in all, special relativity is a 
geometric theory, and thus it is possible to formulate its equations in an invariant form.)  

The spatial axes are transported along the worldline of the origin of the frame in such a way that the angles among them are preserved and that they continue normal to the time axis, the 4--velocity, which implies that 
$\partial_{t}\ushort{e}_{(j)}(t)=\dunderline{\mathfrak{B}}^{(Frame)}_{t}\mycdot\ushort{e}_{(j)}(t)
+\Omega^k_{\ j}(t)\ushort{e}_{(k)}(t)$ with $\Omega^{j}_{\ k}=
\Omega^{j k'}\ushort{e}_{(k')}\mycdot \ushort{e}_{(k)}$, $\Omega^{j k}$ an skew--symmetric $3\times 3$ matrix and $t$ the proper time of the frame. 
The tensor $\dunderline{\mathfrak{B}}^{(Frame)}=\ushort{U}^{(Frame)}\wedge\frac{d}{dt}\ushort{U}^{(Frame)}$ is the FW transport generator associated to the frame, while the tensor 
$\dunderline{\mathfrak{B}}$ that appears in Eq.~\eqref{eq:genevol} is the one associated to the particle. 

The spin vector is to be decomposed not in the basis $\ushort{e}_{(\alpha)}(t)$ at the origin, but in another basis 
$\ushort{e}_{(\alpha)}(\bs{r}_t,t)$ depending on the position of the spin. The two bases differ because in an accelerated reference frame 
a curvature appears. Details of the calculations will be shown in a forthcoming paper \cite{DiLorenzoprep}. 

The components of the spin 4--vector are
$S^\alpha(t)=\ushort{S}(t')\mycdot \ushort{e}^{(\alpha)}(\bs{r}_t,t)$, where 
$\ushort{e}^{(\alpha)}(\bs{r}_t,t)$ are the reciprocal spatial axes of the frame, $\ushort{e}^{(\alpha)}(\bs{r}_t,t)\mycdot \ushort{e}_{(\beta)}(\bs{r}_t,t)=\delta^\alpha_{\ \beta}$.

Thus 
\bal
\partial_{t} S^\alpha(t)
=&\ \frac{1}{\bar{\gamma}_t} \left(\partial_{t'}\ushort{S}_{t'}\right)\mycdot \ushort{e}^{(\alpha)}(\bs{r}_t,t) +\ushort{S}_{t'}\mycdot 
\left(\partial_t\ushort{e}^{(\alpha)}(\bs{r}_t,t)\right)
\label{eq:indexheavy}
\eal
The factor $\bar{\gamma}_t=[\partial t'/\partial t]^{-1}$ is the time dilation, which in an accelerated frame is 
\beq
\bar{\gamma}_t = \frac{1}{\sqrt{(1-\bs{g}_t\mycdot\bs{r}_t)^2-[\bs{v}_t-\bs{\Omega}_t\times\bs{r}_t]^2}},  
\eeq
$\bs{g}$ being the gravitational field at the origin of the frame due to the equivalence principle, \ie\ minus the proper acceleration, 
$\bs{\Omega}$ the Coriolis gravi--magnetic field, \ie\ the angular velocity of the axes of a FW frame as measured by the accelerated frame, and $\bs{r}_t$ the position of the particle. 
We shall define the scalar potential $\theta_t=1-\bs{g}_t\mycdot\bs{r}_t$ and the vector potential 
$\bs{w}_t=\bs{\Omega}_t\times\bs{r}_t$.

The additional terms due to the second addend on the rhs of Eq.~\eqref{eq:indexheavy} lead to a redefinition of the time derivative as a tensor operator, 
\beq
D^\alpha_{\ \beta}=\delta^\alpha_{\ \beta}\partial_t - \left(\frac{d}{dt}\ushort{e}^{(\alpha)}(\bs{r}_t,t)\right)\mycdot \ushort{e}_{(\beta)}(\bs{r}_t,t) ,
\eeq
so that the laws of physics are formally invariant even when written in terms of components. 
By contrast, the formulation of the preceding subsection in terms of 
abstract vectors and tensors does not require any redefinition of derivatives.

After some calculations, 
we find that the spatial components of the spin 4--vector obey 
\bal
\label{eq:accframe}
\partial_t \bs{S}_t =&\ 
\left[\frac{\lambda}{\bar{\gamma}_t}\mathcal{E}_t+\mathcal{K}_t+\mathcal{N}_t\right]
\mycdot\bs{S}_t\ ,
\eal
where the matrices $\mathcal{E}$, $\mathcal{K}$,  and $\mathcal{N}$ 
represent, respectively,  the electromagnetic, the kinematic, 
and the non inertial contribution, 
\bsubs%
\label{eq:matrices}
\bal
\label{eq:matrintr}
\mathcal{E}_t&= -\left[\id_3+\bar{\gamma}^2_t\bs{v}_t\bs{v}^\tr_t-\frac{1}{\theta_t^2}\bs{w}_t\bs{w}^\tr_t\right]\mycdot
\left(\bs{B}_t-\bs{v}^\mathrm{c}_t{\vecp}\bs{E}_t\right)
\mycdot\bs{\mathcal{J}}
,\\
\label{eq:matrkin}
\mathcal{K}_t&=\bs{v}_t
\bs{A}^{\mathrm{sim},\tr}_t\mycdot\left[\id_3-\bs{w}_t\bs{v}^{\mathrm{c},\tr}_t\right]
,\\
\label{eq:matrn}
\mathcal{N}_t&=
\left[\theta_t\bs{g}_t+\partial_t\bs{w}_t
-\frac{1}{\theta_t}
(\partial_t\theta_t+\bs{w}_t\mycdot\bs{g}_t)\bs{w}_t
-\bs{\Omega}_t\times\bs{w}_t\right]
\bs{v}_t^{\mathrm{c},\tr}
+\bs{\Omega}_t\mycdot\bs{\mathcal{J}}
+\frac{1}{\theta_t}
\bs{w}_t\bs{g}_t^\tr
, 
\eal
\esubs%
$\id_3$ being the $3\times 3$ identity and 
$\bs{a}\bs{b}^\tr$ being a matrix written as a dyadic. 
The symbol $\bs{\mathcal{J}}$ denotes the 3--vector having as components the $3\times 3$ generators of spatial rotations, such that for any pair $\bs{a}, \bs{b}$ of 3--vectors 
$\bs{a}\mycdot\bs{\mathcal{J}}\, \bs{b}=\bs{a}\vecp\bs{b}$ and $\bs{b}^\tr\bs{a}\mycdot\bs{\mathcal{J}} =(\bs{b}\vecp\bs{a})^\tr$. 
We introduced the simultaneous 3--acceleration $\bs{A}^\mathrm{sim}=\bs{A}-A^0\bs{v}$, $A^0=\bs{v}^\mathrm{c}\mycdot\bs{A}$, and the  
conjugate velocity 
\beq
\bs{v}^\mathrm{c}_t=\frac{\bs{v}_t-\bs{w}_t}{\theta_t^2-\bs{w}_t^2+\bs{w}_t\mycdot\bs{v}_t} .
\eeq

The time component satisfies identically $S^0(t) = \bs{v}^\mathrm{c}_t\mycdot\bs{S}_t$.

In order to identify correctly the electric and magnetic field in an accelerated frame, some care must be taken because the metric is not flat. One should use the fully covariant representation
\beq
 \ushort{e}_{(\alpha)}(\bs{r}_t,t)\mycdot \dunderline{F}\mycdot \ushort{e}_{(\beta)}(\bs{r}_t,t) = \bp0&-\bs{E}^\tr\\ \bs{E}&-\bs{B}\mycdot\bs{\mathcal{J}}\ep \ .
\eeq

In a nonrotating frame, \ie\ a FW frame, $\bs{\Omega}=0$, and the equivalent gravitational field is Newtonian, as it derives from a scalar potential, hence 
\bal
\label{eq:accfwframe}
\partial_t \bs{S}_t =&\ 
-\frac{\lambda}{\bar{\gamma}_t}\left[\id_3+\bar{\gamma}^2_t\bs{v}_t\bs{v}^\tr_t\right]\mycdot
\left[\left(\bs{B}_t-\frac{\bs{v}_t}{\theta_t^2}{\vecp}\bs{E}_t\right)
\vecp\bs{S}_t\right]+\bs{v}_t
\bs{A}^{\mathrm{sim}}_t\mycdot\bs{S}_t+\frac{\bs{g}_t}{\theta_t}\bs{v}_t\mycdot\bs{S}_t .
\eal
\subsubsection*{Equation in a comoving frame}
In the following, a prime with an asterisk denotes the representation in a comoving frame, while a prime denotes the representation in the FW comoving frame.

In a comoving frame $\bs{v}=\bs{0}$ and $\bs{r}=\bs{0}$, thus 
$\theta=1$, $\bs{w}=\bs{0}$, and Eq.~\eqref{eq:accframe} simplifies to 
\beq
\label{eq:semisimple}
\partial_{t'}\boldsymbol{S}'^*_{t'} = \boldsymbol{S}'^*_{t'}\vecp[\lambda\boldsymbol{B}'^*_{t'} -\bs{\Omega}'^*_{t'}] ,
\eeq
where $\boldsymbol{B}'^*_{t'}$ is the magnetic field in the comoving frame, which is the axial 3D-vector representation of the spatial block of the skew--symmetric tensor 
$\dunderline{\Pi}_{t'}\mycdot\dunderline{F}_{t'}\mycdot\dunderline{\Pi}_{t'}$.

For FW frames $\bs{\Omega}'=\bs{0}$. 
Thus, Eq.~\eqref{eq:genevol} becomes particularly simple in the comoving FW frame: 
\beq
\label{eq:simple}
\partial_{t'}\boldsymbol{S}'_{t'} = \lambda\boldsymbol{S}'_{t'}\vecp\boldsymbol{B}'_{t'}  .
\eeq

The electromagnetic field felt by the particle $\bs{E}', \bs{B}'$ is not given by the standard textbook formulas 
$\bs{E}_\parallel+\gamma(\bs{E}_\perp+\bs{v}\vecp\bs{B})$, $\bs{B}_\parallel+\gamma(\bs{B}_\perp-\bs{v}\vecp\bs{E})$, because 
the Lorentz transformation $L_t$ from the lab frame to the comoving frame is not a simple boost, but includes the Thomas--Wigner rotation. See Eq.~\eqref{eq:restemfield} for an example. 

%
\subsubsection*{Equation in an inertial frame}
In an inertial frame, the axes do not vary. In particular, $\bs{g}=\bs{0}$ and $\bs{\Omega}=\bs{0}$. 
Hence, Eq.~\eqref{eq:accframe} simplifies to 
\bsubs\label{eq:inlab}
\bal
\partial_t \bs{S}_t =&\ 
\mathcal{M}(t)\mycdot\bs{S}_t\ ,
\label{eq:laba}\\
\partial_t \bs{T}_t =&\ -\mathcal{M}^\tr(t)\mycdot\bs{T}_t
\ .
\label{eq:labb}
\eal
\esubs
where the matrix $\mathcal{M}$ is 
\bal
\mathcal{M}(t)&:=-\left[\id_3+\gamma^2_t\bs{v}_t\bs{v}^\tr_t\right]\mycdot
\left[\lambda\gamma^{-1}_t \left(\bs{B}_t-\bs{v}_t{\vecp}\bs{E}_t\right)\mycdot\bs{\mathcal{J}}
-\bs{v}_t\bs{a}^\tr_t\right],
\eal
The time component satisfies identically $S^0(t) = \bs{v}_t\mycdot\bs{S}_t$. 
The terms on the rhs multiplied by $\lambda$ are the direct terms, the remaining term $\left[\id_3+\gamma^2_t\bs{v}_t\bs{v}^\tr_t\right]\mycdot\bs{v}_t\bs{a}^\tr_t=
\gamma^2_t \bs{v}_t\bs{a}^\tr_t$ for $\bs{S}$, or $-\gamma^2_t \bs{a}_t\bs{v}^\tr_t$ for $\bs{T}$,  
is the kinematic term.

\subsection*{Separation of norm and direction in the motion of a vector}
Here we discuss the linear homogeneous equation for an $N$--dimensional vector  
\beq
{\partial_t \bs{P}(t)}= M(t)\bs{P}(t)\ ,
\label{eq:vecmov}
\eeq
with $M(t)$ a given one--parameter family of matrices. 

We are indicating with a bold symbol a column vector. 
A row vector will be indicated as a transpose, \eg\ $\bs{P}^\tr$. 
The row--column product is implied by juxtaposition of the symbols. Thus, $\bs{a}^\tr \bs{b}$ is the scalar $\bs{a}\mycdot\bs{b}$, while $\bs{a}\bs{b}^\tr$ is a matrix; $M\bs{P}$ is a column vector, $\bs{P}^\tr M$ a row vector, 
and  $\bs{P}^\tr M\bs{P}$ is a scalar. 

If $M(t)$ is skew--symmetric, the norm of $\bs{P}(t)$ is constant, thus the equation 
describes a rotation. 
However, how to proceed if $M(t)$ is not skew--symmetric? Can we still individuate a rotation term? 

First, we shall drop from the problem any component which is possibly conserved and does not enter the equations for the other components. 
This happens if $M(t)$ has some rows and the corresponding columns identically equal to zero. Thus, if $M$ can be written 
in the block form $(\begin{smallmatrix}M_0(t)&0\\0&0\end{smallmatrix})$, we shall consider only the block $M_0$ and the corresponding components of 
$\bs{P}$, since the remaining components are constant. 

More generally, if $M(t)$ has a set of $\mu$ common, constant in $t$, left and right eigenvectors $\bs{e}_m$ with 
$0$ eigenvalue, we change coordinates so that the last $\mu$ axes coincide with $\bs{e}_m$. Then, since $\bs{P}(t)\cdot \bs{e}_m=const$, we consider 
only the remaining components of $\bs{P}$, and the corresponding non--zero block of $M$. 

Next, we separate $M(t)$, or its relevant part $M_0(t)$, 
in a symmetric and an skew--symmetric part, $M(t)=\overline{M}(t)+M_\mathrm{a}(t)$. 
The skew--symmetric part gives automatically a contribution perpendicular to $\bs{P}$, hence it provides a rotation. 
The symmetric part, however, also provides a rotation, albeit a nonlinear one, as we shall see in the following. 

We write the vector as $\bs{P}(t)=P(t)\bs{\Hat{P}}(t)$, and we separate the equations 
for the norm $P(t)$ and for the direction $\bs{\Hat{P}}(t)$. By construction, the equation for the direction can be only a rotation. 
By multiplying Eq.~\eqref{eq:vecmov} with $\bs{P}(t)$, we get the equation for the norm, which involves only the symmetric part of $M$, 
\beq\label{eq:modeq}
P(t)\frac{\partial {P}(t)}{\partial t} = P^2(t)\, \bs{\Hat{P}}^\tr(t) \overline{M}(t)\bs{\Hat{P}}(t)\ .
\eeq 
The direction instead satisfies 
\beq
\frac{\partial \bs{\Hat{P}}(t)}{\partial t} = [\overline{M}(t)-\bs{\Hat{P}}^\tr(t) \overline{M}(t) \bs{\Hat{P}}(t)]\bs{\Hat{P}}(t)+M_\mathrm{a}(t)\bs{\Hat{P}}(t)\ .
\eeq
The first term in the rhs can be written 
\beq
[\overline{M}(t)-\bs{\Hat{P}}^\tr(t) \overline{M}(t) \bs{\Hat{P}}(t)]\bs{\Hat{P}}(t)=
A_\mathrm{nl}(t,\Hat{\bs{P}}(t))\bs{\Hat{P}}(t) \ ,
\label{eq:firstterm}
\eeq
where we defined the skew--symmetric matrix 
\beq
A_\mathrm{nl}(t,\Hat{\bs{P}}):=[\overline{M}(t),\bs{\Hat{P}}\bs{\Hat{P}}^\tr]
\ ,
\label{eq:antisymdef}
\eeq 
with $[\ ,\ ]$ the commutator. 
The skew--symmetric matrix $A_\mathrm{nl}(t,\Hat{\bs{P}})$
hence corresponds to another rotation, which is nonlinear, as it depends on the direction $\bs{\Hat{P}}$ itself.

From Eq.~\eqref{eq:modeq}, the square norm can be written after possibly solving for the direction, 
\beq
P^2(t)=P^2(0)\,\exp{[\,2\!\int_0^t dt\, \bs{\Hat{P}}^\tr(t) \overline{M}(t)\bs{\Hat{P}}(t)]}.
\eeq
\subsubsection*{Three dimensional case}
In particular, in three dimensions, to any skew--symmetric matrix it is associated, through the Hodge dual operation, an axial vector which can be identified as an angular velocity. The linear angular velocity is  
\beq
\bs{\omega}^\mathrm{l}(t)\mycdot \bs{\mathcal{J}} =  M_\mathrm{a}(t) \ ,
\eeq
with $\bs{\mathcal{J}}$ the $3\times 3$ angular momentum generators, which form a basis for skew--symmetric matrices. 
The nonlinear angular velocity satisfies  
\beq
\bs{\omega}^\mathrm{nl}(t,\bs{\Hat{P}})\mycdot \bs{\mathcal{J}}
=A_\mathrm{nl}(t,\Hat{\bs{P}}) .
\eeq
Explicitly, 
\beq
\bs{\omega}^\mathrm{nl}(t,\bs{\Hat{P}})=\bs{\Hat{P}}\vecp \left(\overline{M}(t)\bs{\Hat{P}}\right)\ .
\eeq
Finally, in three dimensions, the equations for the direction can be separated as
\beq\label{eq:direq}
\frac{\partial \bs{\Hat{P}}(t)}{\partial t} =[\bs{\omega}^\mathrm{l}(t)+ \bs{\omega}^\mathrm{nl}(t,\bs{\Hat{P}}(t))
]\vecp\bs{\Hat{P}}(t).
\eeq
%
\subsubsection*{Two--dimensional case}
Suppose that the $Z$ component of the vector is conserved, and that it does not enter in the equation for the other two components. 
This case is relevant because often times the trajectory is planar, and the off-plane component of the spin is always conserved. 
Let $M_\RN{2}$ the $2\times 2$ nonzero block of $M$. 
Let $\bs{P}_\parallel=\{\bs{P}_\RN{2},0\}$ the component of $\bs{P}$ on the $XY$ plane with $\bs{P}_\RN{2}$ the 2--component vector consisting in the first two components of $\bs{P}$. 
We have an effective two--dimensional equation 
\beq
\partial_t \bs{P}_\RN{2}(t)=M_\RN{2}(t) \bs{P}_\RN{2}(t) . 
\eeq
We may proceed as in the general case, and apply Eq.~\eqref{eq:antisymdef} to $\bs{P}_\RN{2}(t)$. 
Since there is only one skew--symmetric matrix in 2D, up to a constant, $A_Z=(\begin{smallmatrix}0&-1\\1&0\end{smallmatrix})$, the commutator yields 
\beq
A_\mathrm{nl}(t,\Hat{\bs{P}}_\parallel):=[\overline{M}_\RN{2}(t),\bs{\Hat{P}}_\RN{2}\bs{\Hat{P}}_\RN{2}^\tr]=\omega^\mathrm{nl}(t,\bs{\Hat{P}}_\parallel(t)) A_Z, 
\eeq
where $\omega^\mathrm{nl}(t,\bs{\Hat{P}}_\parallel)=\bs{\Hat{k}}\mycdot[\bs{\Hat{P}}_\parallel(t)\vecp(\overline{M}(t)\bs{\Hat{P}}_\parallel(t))]$, 
while $\bs{\Hat{P}}_\parallel(t)$ is the unit vector of the component of $\bs{P}$ parallel to the $XY$ plane, \ie\ if $\bs{P}=\{X,Y,Z\}$,  
$\bs{\Hat{P}}_\parallel=\{X,Y,0\}/\sqrt{X^2+Y^2}$. 
Better, in invariant form $|\bs{P}_\parallel|=|\bs{P}\vecp \bs{\Hat{k}}|$, 
$\bs{\Hat{P}}_\parallel =(\bs{P}-\bs{\Hat{k}}\mycdot\bs{P}\, \bs{\Hat{k}})/|\bs{P}_\parallel|$.

The nonlinear rotation rate, as a vector in 3D, is thus 
\beq
\bs{\omega}^\mathrm{nl}(t,\bs{\Hat{P}}) = \frac{\{\bs{\Hat{P}}\vecp[\overline{M}(t)\bs{\Hat{P}}]\}\mycdot\bs{\Hat{k}}\, \bs{\Hat{k}}}{|\bs{\Hat{P}}\times\bs{\Hat{k}}|^2}
\eeq

Had we not eliminated from the equations the constant component $Z$, and had we instead applied the result for the 3D case, we would get in--plane terms in $\bs{\omega}^\mathrm{nl}$, which would change 
the $Z$ component of the unit vector $\bs{\Hat{P}}$ in such a way as to compensate the change in the norm $P$. 
\subsubsection*{Application to the spin precession}
In our case, in absence of external fields, the equation for the simultaneity--corrected spin in an inertial frame is 
\beq
\frac{\partial}{\partial t} \bs{T}(t) = -\gamma^2(t)\bs{a}(t) \bs{v}(t)\mycdot\bs{T}(t), 
\eeq
hence the matrix is 
\beq
M(t) = -\gamma^2(t)\bs{a}(t) \bs{v}^\tr(t) .
\eeq
The symmetric part is 
\beq
\overline{M}(t)=-\frac{1}{2}\gamma^2(t)\frac{d}{dt}\bs{v}(t)\bs{v}^\tr(t)\ .
\eeq
Therefore, the nonlinear part of the precession (since a spin represents a rotation, the rotation of a spin is called a precession), 
\bal
\bs{\omega}^\mathrm{nl}(t,\bs{\Hat{T}})=&\ -\frac{\gamma^2(t)}{2}\bs{\Hat{T}}\vecp\left\{\left[\frac{d}{dt}\bs{v}(t) \bs{v}^\tr(t)\right]\bs{\Hat{T}}\right\} \ .
\eal

However, if the motion is planar, the component of the spin normal to the plane of motion in the lab frame is conserved. Thus, we should use the 2D result, which yields a nonlinear rotation rate 
\bal
\bs{\omega}^\mathrm{nl}(t,\bs{\Hat{T}})=&\ -\frac{\gamma^2(t)}{2}\left(\bs{\Hat{T}}_\parallel\vecp\left\{\left[\frac{d}{dt}\bs{v}(t) \bs{v}^\tr(t)\right]\bs{\Hat{T}}_\parallel\right\}\right)\mycdot\bs{\Hat{k}}\, \bs{\Hat{k}}\ .
\eal

The skew--symmetric term is
\beq
M_\mathrm{a}=-\frac{1}{2}\gamma^2(t)[\bs{a}(t)\bs{v}^\tr(t)-\bs{v}(t)\bs{a}^\tr(t)]\ ,
\eeq
corresponding to the Thomas precession
\beq
\bs{\omega}^\mathrm{Th}(t)= -\frac{\gamma^2(t)}{2}\bs{v}(t)\vecp\bs{a}(t)  \ .
\eeq
For the conventional spin $\bs{S}$ the same linear term holds, while the non--linear term has the opposite sign.
%

%

\section*{Acknowledgements}
This work was performed as part of the Brazilian Instituto Nacional de Ci\^{e}ncia e
Tecnologia para a Informa\c{c}\~{a}o Qu\^{a}ntica (INCT--IQ), it
was supported by CNPq, Conselho Nacional de Desenvolvimento Cient\'{\i}fico e Tecnol\'{o}gico, proc. 312466/2017-0, and by FAPEMIG, Funda\c{c}\~{a}o de Amparo \`{a} Pesquisa de Minas Gerais, PPM--00607--16 and PPM--00180--18.
\section*{Author contributions statement}
As the sole author, A.D.L. made the research, secured the funding, and wrote the manuscript.

\section*{Competing interests}
The author declares no competing interests.

\section*{Data availability}
All data generated or analysed during this study are included in this published article (and its Supplementary Information files).

\appendix

%
\section{An overview of the notation and known results\label{app:overview}} 
\subsection{Notation}
We will use an underscore for abstract vectors in Minkowski space, while an arrow will indicate a contravariant representation of 
the vector in some reference frame, which we define below. Greek labels range in $\{0,1,2,3\}$, while Latin labels range in $\{1,2,3\}$. 
The convention of summing over repeated indexes is used, so that $u^j v_j=\sum_{j=1}^3 u^j v_j$, $u^\alpha v_\alpha=\sum_{\alpha=0}^3 u^\alpha v_\alpha$.  
However, as soon as possible, we shall replace sums over components with dot products, for ease of readability.  
The signature of the metric is chosen as ${-}{+}{+}{+}$.

Observers are represented by worldlines in Minkowski spacetime, usually indicated by a letter $O$, primed or not. The worldline can be parametrized with its natural parameter, the proper time $\tau_O$ or $t$ for short.  
A reference frame $\mathfrak{S}$ consists in a worldline $O$, along with a set of 4 vectors, $\ushort{e}_{(\alpha)}(t)$, such that the first one 
$\ushort{e}_{(0)}(t)$ is the unit tangent vector to the worldline ($\ushort{e}_{(0)}(t)\mycdot\ushort{e}_{(0)}(t)=-1$), while the other three 
$\ushort{e}_{(j)}(t)$ are orthogonal to $\ushort{e}_{(0)}(t)$. The latter three are the spatial axes of the reference frame. 
While they need not be mutually orthogonal, their relative angles are required to be fixed, so that the metric, defined by 
$g_{\alpha\beta}:=\ushort{e}_{(\alpha)}(t)\mycdot\ushort{e}_{(\beta)}(t)$, is constant along the worldline, and furthermore has 
the block representation $g=\left(\begin{smallmatrix}-1&\boldsymbol{0}\\\boldsymbol{0}&\mathcal{G}\end{smallmatrix}\right)$, with $\mathcal{G}$ a constant 
$3\times 3$ matrix. 

The metric allows to define the reciprocal vectors 
$\ushort{e}^{(\alpha)}(t)=(g^{-1})^{\alpha\beta}\ushort{e}_{(\beta)}(t)$, with $g^{-1}$ the inverse of $g$. 
Thus $\ushort{e}^{(0)}=-\ushort{e}_{(0)}$. 
One may want to choose $\mathcal{G}$ to be the $3\times 3$ identity, by picking an orthonormal basis, in which case 
$\ushort{e}^{(j)}=\ushort{e}_{(j)}$. The corresponding matrix is then the Minkowski metric $g=g^{-1}=\eta$. 
This choice, however is not cogent: the laws of Nature are indifferent to the fact that we find easier expressing them in an orthonormal basis.   
The orthogonality between the time axis $\ushort{e}_{(0)}$ and the spatial axes $\ushort{e}_{(j)}$, on the other hand, is fundamental, not an arbitrary choice. 
\subsection{Contravariant and covariant vector representations}
In a given basis, the components of a vector transported along the same worldline, $\ushort{V}(t)=V^\alpha(t) \ushort{e}_{(\alpha)}(t)$ are obtained by multiplying times the reciprocal basis 
$V^\alpha(t)=\ushort{V}(t)\mycdot\ushort{e}^{(\alpha)}(t)$. 
These components, collected in a column vector, will be denoted by the symbol $\vec{V}$, which is the representation of $\ushort{V}$ in the reference frame defined by the basis $\ushort{e}_{(\alpha)}$. For increased readability, a column vector will be 
written as a horizontal list in braces $\{V^0,V^1,V^2,V^3\}$. One may want to write the components in the reciprocal basis, which will be denoted by a lower index, 
$V_\alpha:=\ushort{V}\mycdot\ushort{e}_{(\alpha)}$. In this case we shall enclose the list in round parentheses, $(V_0,V_1,V_2,V_3)$, 
consider the vector as a row vector, and indicate it as $\vec{V}^\conj$. Given the metric representation $g$, $\vec{V}^\conj=[g\mycdot \vec{V}]^\tr$. Furthermore $V_0=-V^0$, while the spatial components are $(V_1,V_2,V_3)=[\mathcal{G}\cdot\bs{V}]^\tr$. In case one is using an orthonormal spatial basis, $V_j=V^j$. 

In a given representation, we shall indicate  column 3--vectors by boldface letters, $\boldsymbol{v}, \boldsymbol{a}$, while the row vectors will be denoted by a boldface symbol with a conjugation, $\boldsymbol{v}^\conj, \boldsymbol{a}^\conj$. 
In particular, for the Minkowski metric, since $\mathcal{G}=\id_3$, instead of the conjugation symbol we shall use the conventional transpose, $\bs{v}^\tr$, \etc.
We will also use the notation $\bs{v}\bs{a}^\conj$ to indicate the direct product, a $3\times 3$ matrix having in the $j$--th row and $k$--th column the value 
$v^j a_k$.

Vectors per se are invariant objects. Their components however, transform with the basis. The vector representations  
$\vec{U}$ and $\vec{U}^\conj$ are called contravariant and covariant vectors, resp. A stricter term would be \emph{contravariant and covariant vector representations}. 

The basis in its own representation is simply the canonical basis: $\vec{e}_{(0)}=\{1,0,0,0\}$, $\vec{e}_{(1)}=\{0,1,0,0\}$, etc. 
If one considers another basis $\ushort{f}_{(\alpha)}$, the former basis can be represented in it as, \eg, $\vec{e}^{\,[f]}_{(0)}=\{e_{(0)}^0,e_{(0)}^1,e_{(0)}^2,e_{(0)}^3\}$. Here, a label distinguishing a vector is indicated in parentheses, while an index with no parentheses around it denotes a component in a given basis. 
\subsection{Classification of reference frames\label{app:fw}}
Inertial reference frames have a straight worldline $O$, which makes $\ushort{e}_{(0)}$ a constant vector, 
and constant spatial axes $\ushort{e}_{(j)}(t)=\ushort{e}_{(j)}(0)$. A frame with constant $\ushort{e}_{(0)}$ but time-varying 
$\ushort{e}_{(j)}(t)$ is a purely rotating frame. A frame with varying time axis $\ushort{e}_{(0)}(t)$ is an accelerated frame. 

Among the accelerated frames sharing the same worldline $O$, there is a special one, the FW frame: 
The normality condition $\ushort{e}_{(0)}(t)\mycdot \ushort{e}_{(j)}(t)=0$ implies that for an accelerated frame the spatial axes must also vary in time. 
If we require that their time derivatives $d \ushort{e}_{(j)}(t)/dt$ are parallel to $\ushort{e}_{(0)}(t)$, no noninertial rotations are induced in the accelerated frame. A reference frame satisfying this condition is called a FW frame (other texts refer to it as Fermi coordinates). 

Consider indeed, at a given time, a purely spatial vector representing the position of a particle $E$ at time $t$ relative to the observer: $E(t)-O(t)$, which means that the vector is perpendicular to the wordline $O$. 
The spatial coordinates are given by $X^j(t) = [E(t)-O(t)]\mycdot \ushort{e}^{(j)}(t)$. At a later time, the new coordinates of the particle are, to first order, 
\bal
X^j(t+\delta t) =&\ [E(t+\delta t)-O(t+\delta t)] \mycdot \ushort{e}^{(j)}(t+\delta t) 
\nonumber
\\
\simeq&\ 
 \left\{E(t)-O(t)+\delta t\left[\gamma^{-1}_{EO}(t)\ushort{U}_E(t)  - \ushort{U}_O(t)\right]\right\}\mycdot [\ushort{e}^{(j)}(t)+\delta t \frac{d}{dt} \ushort{e}^{(j)}(t)]
\nonumber
\\
\simeq&\  
X^j(t) + \delta t v^j_{E|O}(t) + \delta t [E(t)-O(t)]\mycdot  \frac{d}{dt} \ushort{e}^{(j)}(t) ,
\eal
with $\ushort{U}_E$ the 4--velocity of the particle $E$, $\ushort{U}_O(t)=\ushort{e}_{(0)}(t)$ the 4--velocity of the observer $O$, 
and $\gamma_{EO}$ the Lorentz dilation factor $[1-v^2_{E|O}(t)]^{-1/2}$ for the velocity of the particle $E$ relative to $O$.  
If the derivatives $\frac{d}{dt} \ushort{e}^{(j)}(t)$ contain spatial components, the last term in the rhs 
of the last equality will contribute a noninertial rotation to the movement of the particle as described in the accelerated reference frame. If instead they are parallel to $\ushort{U}_O$, the product vanishes, because  
$E(t)-O(t)$ is normal to the trajectory. 

This latter condition is the relativistic analog of the classical case, where, given an accelerated origin $O$, we may consider, among all possible 
comoving frames, one in which the axes are fixed relative to an inertial frame. In special relativity, however, one cannot impose that the spatial axes 
remain fixed $d\ushort{e}^{(j)}(t)/dt = 0$, because they must be kept orthogonal to the time axis, which in an accelerated frame changes direction in 
Minkowski space. However, by requiring that the derivatives are parallel 
to the 4--velocity, one ensures that no Coriolis rotations appear in the motion of $E$, hence we can say that the reference frame so defined corresponds 
to the non--relativistic non--rotating frame.
\subsection{Transformations between two frames}
Upon a change of basis, the representation of a vector $\ushort{V}$ transforms according to 
\beq
V^{[f],\alpha} =\ushort{V}\mycdot \ushort{f}^{(\alpha)} = \ushort{V}\mycdot \ushort{e}^{(\beta)} \ushort{e}_{(\beta)}\mycdot\ushort{f}^{(\alpha)} = 
T^\alpha_{\ \beta} V^{[e],\beta} ,
\eeq
or, more compactly,  
\beq
\vec{V}^{[f]} = T\mycdot \vec{V}^{[e]} \ .
\label{eq:basistransf}
\eeq
The matrix $T^\alpha_{\ \beta}:=\ushort{f}^{(\alpha)}\mycdot\ushort{e}_{(\beta)}$ provides the change of vectors applied at the origin. 
The upper index indicates the rows, and the lower index the columns, so that the dot ``\(\mycdot\)'' symbol in Eq.~\eqref{eq:basistransf} 
indicates the ordinary row-column product.

We are overloading the same symbol with different meanings, which, however, are clear from the objects to which it is applied: 
$\ushort{A}\mycdot\ushort{B}$ is the product in Minkowski space, $\vec{A}^\tr\mycdot\vec{B}$ is the row--column product, and the dot between two column vectors $\vec{A}\mycdot\vec{B} = (\vec{A}^\tr\mycdot g)\mycdot \vec{B}$ is the product mediated by the metric. We shall avoid the last usage, however, limiting it to the 3--dimensional subspace when a non--orthonormal basis is being used, so that we will write $\bs{v}\mycdot\bs{a}$ 
instead of $\bs{v}^\conj\mycdot\bs{a}$ for the product $v_k a^k = v^j \mathcal{G}_{jk} a^k$. 

For brevity, since we shall deal only with a handful of bases, rather than indicating with $\vec{V}^{[e]},\vec{V}^{[f]}$, etc., which basis a 4--vector is being represented in, we shall use one or more apexes, \eg\, $\vec{V}, \vec{V}'$. 

\subsection{Lorentz transformations}
In particular, a transformation between two frames both of which are using orthonormal bases is said to be a Lorentz transformation, which we will denote as $L$. 
In this case, the inverse satisfies 
\bal
(L^{-1})^\alpha_{\ \beta}:=&\ \ushort{e}^{(\alpha)}\mycdot\ushort{f}_{(\beta)} =\eta^{\alpha\alpha'}\eta_{\beta\beta'}\ushort{e}_{(\alpha')}\mycdot\ushort{f}^{(\beta')} 
\nonumber
\\
=&\ 
\begin{cases}
&\phantom{-}(L^\tr)^\alpha_{\ \beta} , \text{if } \alpha,\beta \in \{1,2,3\}, \text{ or if } \alpha=\beta=0 \\
&-(L^\tr)^\alpha_{\ \beta} , \text{if } \alpha \in \{1,2,3\}\text{ and }\beta=0, \text{ or vice versa} 
\end{cases} .
\label{eq:justlorentz}
\eal
Let $\bs{V}^\tr=-(L^0_{\ 1},L^0_{\ 2},L^0_{\ 3})$ the row 3--vector,  $\bs{V}'=\{L^1_{\ 0},L^2_{\ 0},L^3_{\ 0}\}$ the column 3--vector, and $\mathcal{L}$ the $3\times 3$ spatial block of the transformation matrix, which thus has the block form
\beq
L=\ \begin{pmatrix}L^0_{\ 0}&-\bs{V}^{\tr}\\
\bs{V}'&\mathcal{L}
\end{pmatrix}
.
\label{eq:blocklorentz}
\eeq
The conditions on the inverse Lorentz transformation can be written in the block matrix form 
\beq
L^{-1}=
 \begin{pmatrix}L^0_{\ 0}&-\bs{V}'^{\tr}\\\
\bs{V}&\mathcal{L}^\tr
\end{pmatrix}.
\eeq
By equalling $L^{-1} \mycdot L$ and $L \mycdot L^{-1}$ to the identity, 
we find the conditions: 
\bsubs
\bal
[L^0_{\ 0}]^2=&\ 1+\bs{V}\mycdot\bs{V}=1+\bs{V}'\mycdot\bs{V}' ,
\label{eq:lor00} 
\\
\mathcal{L}\mycdot\bs{V}+L^0_{\ 0} \bs{V}'=&\ \mathcal{L}^\tr\mycdot\bs{V}'+L^0_{\ 0} \bs{V}=0, 
\label{eq:lor01} 
\\
\mathcal{L}\mycdot\mathcal{L}^\tr-\bs{V}'\bs{V}'^{\tr}=&\ \mathcal{L}^\tr\mycdot\mathcal{L}-\bs{V}\bs{V}^{\tr}=\id_3, 
\label{eq:lor11} 
\eal
\label{eq:lor}%
\esubs
or, in components 
\bsubs
\label{eq:lorentzcondition}
\bal
-L^\alpha_{\ 0} L^\beta_{\ 0}+\sum_j L^\alpha_{\ j} L^\beta_{\ j} =& \eta^{\alpha\beta} , \\
-L^0_{\ \alpha} L_{\ \beta}^0+\sum_j L^j_{\ \alpha} L_{\ \beta}^j =& \eta_{\alpha\beta} .
\eal
\esubs

In Ref.~\cite{DiLorentz}, we discuss the classification of Lorentz transformations based on their eigenvalues, and on which axes, if any, they leave unchanged.  
In particular, here, two special cases are of interest: spatial rotations, which leave the time axis and a spatial axis unchanged, and boosts, which leave two spatial 
axes unchanged. Boosts are characterized by the property that they leave unchanged two spatial axes orthogonal to the relative velocity, which also implies 
that, if the 
velocity of the final frame relative to the initial frame is $\bs{v}$,  the reciprocal 
velocity $\bs{v}'$ is $-\bs{v}$. This latter property, however, is not unique to boosts, as it is shared with 4-screws \cite{Synge}, transformations consisting in  
a boost accompanied by a rotation about the axis of the boost. 
Spatial rotations, on the other hand, are characterized by a vector angle $\bs{\phi}$ encoding the direction of the rotation and the angle. 
Boosts are symmetric matrices, rotations are orthogonal matrices 
\beq
B[\bs{v}]=\bp \gamma&-\gamma \bs{v}^\tr\\-\gamma\bs{v}&\id_3+\frac{\gamma^2}{\gamma+1}\bs{v}\bs{v}^\tr\ep,  \qquad R[\bs{\phi}]=
\bp1&\bs{0}^\tr\\\bs{0}&\mathcal{R}[\bs{\phi}]\ep, 
\label{eq:br}
\eeq
where $\mathcal{R}[\bs{\phi}]=\exp[\bs{\phi}\mycdot\bs{\mathcal{J}}]$, with $\bs{\mathcal{J}}$ the vector having as components the $3\times 3$ representations of the 
angular momentum 
\beq
\mathcal{J}_1=\bp0&0&0\\0&0&-1\\0&1&0\ep , \quad \mathcal{J}_2=\bp0&0&1\\0&0&0\\-1&0&0\ep, \quad \mathcal{J}_3=\bp0&-1&0\\1&0&0\\0&0&0\ep .
\eeq 

It is well known that a general proper orthochronous Lorentz transformation $L$ can be decomposed uniquely as the product of a boost and a proper rotation: 
\beq
L = R'[\bs{\phi}]\ \mycdot B[\boldsymbol{v}] = B[\boldsymbol{v}^*]\mycdot R[\bs{\phi}] ,
\label{eq:decom}
\eeq
where $\boldsymbol{v}^*=\mathcal{R}[\bs{\phi}] \mycdot \boldsymbol{v}$ is obtained by the rotation of the 3--vector $\boldsymbol{v}$ applying the spatial submatrix 
$\mathcal{R}$ of $R$.

The solutions to Eq.~\eqref{eq:decom} are built as follows:  
For the decomposition $L=R'B$
\begin{itemize}
\item Take the first row of $L$ in Eq.~\eqref{eq:blocklorentz}, input it as the first row of  $B$ in Eq.~\eqref{eq:br}, and complete the entries of $B$.  
For orthochronous transformations $L^0_{\ 0}>0$, hence this step is always possible. 
In components 
\beq
B^\alpha_{\ \beta}[\bs{v}]=\eta^\alpha_{\ \beta}+\frac{(L^0_{\ \alpha}+\delta^0_{\ \alpha}) (L^0_{\ \beta}+\delta^0_{\ \beta})}{L^0_{\ 0}+1}.
\eeq
\item Consider the $3\times 3$ matrix 
\beq
\mathcal{R}'=\mathcal{L}+\frac{\bs{V}'\bs{V}^\tr}{L^0_0+1} . 
\eeq
A straightforward verification using Eqs.~\eqref{eq:lor} shows that it is an orthogonal matrix. It defines the left rotation $R'$.
In components
\beq\label{eq:rot}
\mathcal{R}'^j_{\ k}=L^j_{\ k}-\frac{L^j_{\ 0} L^0_{\ k}}{L^0_{\ 0}+1} .
\eeq
\end{itemize}
For the second decomposition, analogously, one takes the first column of $L$  in Eq.~\eqref{eq:blocklorentz}, inputs it as the first column of 
$B[\bs{v}^*]$, and completes the entries according to Eq.~\eqref{eq:br}. 
In components 
\beq
B^\alpha_{\ \beta}[\bs{v}^*]=\eta^\alpha_{\ \beta}+\frac{(L^\alpha_{\ 0}+\delta^\alpha_{\ 0}) (L^\beta_{\ 0}+\delta^\beta_{\ 0})}{L^0_{\ 0}+1}.
\eeq
The rotation matrix $R$ is the same as $R'$ in the previous decomposition. 
\subsubsection{Interpretation of the polar decompositions}
The decomposition of the transformation between the frame $\mathfrak{S}$ and $\mathfrak{S}'$, $L=R'B$, 
means that the boost $B$ gives the transformation from the frame $\mathfrak{S}$ to a frame $\mathfrak{S}'^*$ moving 
with velocity $\bs{v}$ relative to $\mathfrak{S}$ and hence at rest in $\mathfrak{S}'$, and relative to which $\mathfrak{S}$ moves with velocity $\bs{v}'^*=-\bs{v}$. 
The rotation $R'$ represents a rotation in the boosted frame, hence the prime. The matrix $R'$ transforms the representation of a 4--vector in 
$\mathfrak{S}'^*$ into its representation in $\mathfrak{S}'$, \ie\ it describes how the axes of 
$\mathfrak{S}'^*$, obtained by a simple boost from $\mathfrak{S}$, are rotated relative to to the axes of the 
frame $\mathfrak{S}'$.
 
The alternative decomposition $L=BR$ has a different meaning. First one applies a rotation $R$ in the frame $\mathfrak{S}$, thus representing 
a 4--vector in a frame $\mathfrak{S}^*$ at rest relative to $\mathfrak{S}$, but with the axes rotated. 
Then one applies a boost in this rotated frame, yielding the representation of the 4--vector in the final frame 
$\mathfrak{S}'$. 
The inverse matrix $R^ {-1}$ thus describes the rotation of $\mathfrak{S}^*$ relative to $\mathfrak{S}$. 
The fact that the frames are connected by a boost means that, since $\bs{v}^*=\mathcal{R}\bs{v}$ is the velocity of $\mathfrak{S}'$ relative to $\mathfrak{S}^*$, the velocity of $\mathfrak{S}^*$, and hence also of 
$\mathfrak{S}$, relative to $\mathfrak{S}'$ is 
$\bs{v}'=-\bs{v}^*$. 
For this reason, we denoted the first column of the Lorentz transformation \eqref{eq:blocklorentz} with the sign opposite to the first row: 
the column $\bs{V}'=\gamma'\bs{v}'$ yields directly the velocity of the original frame relative to the final frame. 

Formally, the rotation matrices are the same $R=R'$. However, they represent different operations, 
hence we used a prime to distinguish them.  
\subsubsection{Minimal rotation \label{subsub:minimalrot}}
While the rotation $\mathcal{R}$ takes $\bs{v}$ to $\bs{v}^*$, in general, the rotation $\mathcal{R}$ is not minimal, \ie\ it is not a rotation about the axis 
$\bs{v}\times\bs{v}^*$. 
In Minkowski space a rotation is no longer characterized by an axis and an angle, but rather by a plane and an angle. It seems appropriate to adapt 
to this characteristic when describing 3D rotations as well. 
In this case, by specifying any two 
unit vectors connected through $\bs{\Hat{n}} = \mathcal{R}\bs{\Hat{m}}$, and requiring that the rotation is in the plane determined by them, we 
arrive to the unique proper 3D--rotation satisfying these conditions, 
\beq
\mathcal{R}(\bs{\Hat{m}},\bs{\Hat{n}})=\id_3 
- \frac{1}{1+\bs{\Hat{m}}\mycdot\bs{\Hat{n}}}
[\bs{\Hat{m}}+\bs{\Hat{n}}][\bs{\Hat{m}}^\tr+\bs{\Hat{n}}^{\tr}]+2\bs{\Hat{n}} \bs{\Hat{m}}^\tr \ .
\label{eq:rotobs}
\eeq
The rotation reduces to the identity when $\bs{\Hat{m}}=\bs{\Hat{n}}$, and it is, naturally, indefinite for $\bs{\Hat{m}}=-\bs{\Hat{n}}$, when a $\tfrac{0}{0}$ 
singularity appears. 

In this case, indeed, writing $\bs{\Hat{n}}=-\cos(\epsilon) \bs{\Hat{m}}+\sin(\epsilon)[\cos(\phi_\epsilon)\bs{\Hat{k}}+\sin(\phi_\epsilon)\bs{\Hat{l}}]$, with $\bs{\Hat{k}}, \bs{\Hat{l}}, \bs{\Hat{m}}$ 
an orthogonal triad, we see that there is no definite limit unless $\lim\limits_{\epsilon\to 0} \phi_\epsilon$ exists. 

The latter case means that $\bs{\Hat{n}}$ tends to $-\bs{\Hat{m}}$ on a fixed plane. 
Without loss of generality, let us say that $\lim\limits_{\epsilon\to 0} \phi_\epsilon=0$. 
Then taking the limit $\epsilon\to 0$, yields
\beq
\mathcal{R}\to \id - 2\bs{\Hat{k}}\bs{\Hat{k}}^\tr -2\bs{\Hat{m}}\bs{\Hat{m}}^\tr \ ,
\eeq
which is a proper rotation of an angle $\pi$ about the axis orthogonal to $\bs{\Hat{m}}$ and $\bs{\Hat{k}}$. 
%
\subsection{Thomas--Wigner's rotation and Stapp's formula}
We apply the result about polar decomposition \eqref{eq:decom} to the combination of two boosts, 
\beq
L=B[\boldsymbol{u'}] B[\boldsymbol{u}] .
\eeq
We are using a prime on the second velocity because when applied to a vector representation $\vec{C}$, the first boost will transform a vector from its representation in a frame $\mathfrak{S}$ to its representation $\vec{C}'$ in a new frame $\mathfrak{S}'$, moving with velocity $\boldsymbol{u}$ relative to $\mathfrak{S}$. 
The second boost will represent $\vec{C}'$ in another frame $\mathfrak{S}''$, moving with velocity $\bs{u}'$ relative to $\mathfrak{S}'$. 
The product of two boosts, as described in my manuscript \cite{DiLorentz}, individuates a special subset of Lorentz transformation, called planar transformations, which preserve the spatial axis orthogonal to the velocities. 

It is convenient to work not with Newtonian velocities, but with the celerities $\bs{U}=\gamma_u \bs{u}$, $\bs{U}'=\gamma_{u'} \bs{u}'$.  
(It is not uncommon to read that $\gamma \bs{v}$ is the proper velocity. However, a proper quantity is one measured in a frame where a particle is momentarily at rest. Thus the proper velocity is always zero. We would like to use the term relativistic velocity for $\gamma\bs{v}$, but it is easily confused with relativistic speed meaning a speed close to that of light. 
To avoid this misnomer, we refer to $\gamma \bs{v}$ with the other attested name in the literature, celerity.) 
Then 
\bal
L =&\  
\begin{pmatrix}
\gamma_{u'}&-\bs{U}'^\tr\\-\bs{U}'&\id_3+\frac{\bs{U}'\bs{U}'^\tr}{\gamma_{u'}+1}
\end{pmatrix}
\begin{pmatrix}
\gamma_{u}&-\bs{U}^\tr\\-\bs{U}&\id_3
+\frac{\bs{U}\bs{U}^\tr}{\gamma_{u}+1}
\end{pmatrix}
\nonumber\\
=&\ \begin{pmatrix}
\gamma_{u'}\gamma_{u}+\bs{U}'\mycdot\bs{U}&-\bs{V}^\tr\\+\bs{V}''&
\id_3+\frac{\bs{U}'\bs{U}'^\tr}{\gamma_{u'}+1}+\frac{\bs{U}\bs{U}^\tr}{\gamma_{u}+1}
+[1+\frac{\bs{U}'\mycdot\bs{U}}{(\gamma_{u'}+1)(\gamma_{u}+1)}]\bs{U}'\bs{U}^\tr
\end{pmatrix}.
\eal
Here, we put 
\bsubs
\bal
\bs{V} =&\  
\bs{U'}+[\gamma_{u'} + \frac{\bs{U}'\mycdot\bs{U}}{\gamma_{u}+1}]\bs{U},\\
\bs{V}'' =&\ -\bs{U}-[\gamma_{u} + \frac{\bs{U}'\mycdot\bs{U}}{\gamma_{u'}+1}]\bs{U}' .
\eal
\label{eq:celerities}%
\esubs
The vector $\bs{V}$ is but the celerity $\bs{U}'$ in the initial reference frame $\mathfrak{S}$, \ie\ 
the celerity of $\mathfrak{S}''$ relative to 
$\mathfrak{S}$. Conversely, $\bs{V}''$ is the celerity of $\mathfrak{S}$ relative to $\mathfrak{S}''$. 

It is impossible to have $\bs{V}=\bs{V}''\neq\bs{0}$. 
Indeed, imposing $\bs{V}=\bs{V}''$ in Eqs.~\eqref{eq:celerities} we find 
\beq
\frac{(\gamma_{u}+1)(\gamma_{u'}+1)+\bs{U}\mycdot\bs{U}'}{\gamma_{u'}+1}\bs{U}' = -\frac{(\gamma_{u}+1)(\gamma_{u'}+1)+\bs{U}\mycdot\bs{U}'}{\gamma_{u}+1}\bs{U} .
\label{eq:v=v''}
\eeq
Since $(\gamma_{u}+1)(\gamma_{u'}+1)+\bs{U}\mycdot\bs{U}'\ge \sqrt{(\gamma_{u}+1)(\gamma_{u'}+1)}[\sqrt{(\gamma_{u}+1)(\gamma_{u'}+1)}-\sqrt{(\gamma_{u}-1)(\gamma_{u'}-1)}]>0$, it follows that Eq.~\eqref{eq:v=v''} can hold only if $\bs{U}=-\bs{U}'$, which implies $\bs{V}=\bs{V}''=\bs{0}$. 

The common Lorentz factor of $\bs{V}$ and $\bs{V}''$ is 
\beq
\gamma_v=\gamma_{u'}\gamma_{u}+\bs{U}'\mycdot\bs{U} .
\eeq
We shall use this relation to replace $\bs{U}'\mycdot\bs{U}$ in terms of the Lorentz factors. 
In particular 
\bsubs
\bal
\bs{V} =&\  
\bs{U'}+\frac{\gamma_{u'} +\gamma_v}{\gamma_{u}+1}\bs{U},\\
\bs{V}'' =&\ -\bs{U}-\frac{\gamma_{u} +\gamma_v}{\gamma_{u'}+1}\bs{U}' .
\eal
\esubs
After replacing $\bs{U}$ and $\bs{U}'$ with $\bs{V}$ and $\bs{V}''$, we have
\bal
L =&\ \begin{pmatrix}
\gamma_{v}&-\bs{V}^\tr\\
\bs{V}''&
\id_3-\frac{(\gamma_{u}+1)(\gamma_{u'}+1)}{(\gamma_v-1)(\gamma_u+\gamma_{u'}+\gamma_v+1)^2}[\bs{V}-\bs{V}'']
[\bs{V}^\tr-\bs{V}''^{\tr}]-\frac{\bs{V}'' \bs{V}^\tr}{\gamma_v-1}
\end{pmatrix}.
\label{eq:b1b2}
\eal

From the polar decomposition 
\beq
L=R\mycdot B[\bs{v}] = 
\begin{pmatrix}
\gamma_{v}&-\bs{V}^\tr\\-\mathcal{R}\mycdot\bs{V}&\mathcal{R}\mycdot(\id_3+\frac{\bs{V}\bs{V}^\tr}{\gamma_{v}+1}) 
\end{pmatrix},
\eeq
we have, by comparison with Eq.~\eqref{eq:b1b2}, 
\beq
\mathcal{R}=\id_3-\frac{(\gamma_{u}+1)(\gamma_{u'}+1)(\gamma_{v}+1)}{(\gamma_u+\gamma_v+\gamma_{u'}+1)^2}
[\bs{\Hat{v}}-\bs{\Hat{v}}'']
[\bs{\Hat{v}}^\tr-\bs{\Hat{v}}''^{\tr}]-2\bs{\Hat{v}}'' \bs{\Hat{v}}^\tr \ ,
\label{eq:rotb1b2}
\eeq
where we divided $\bs{V}$ and $\bs{V}''$ by their common norm $\sqrt{\gamma_v^2-1}$, introducing the corresponding 
unit vectors $\bs{\Hat{v}}$ and $\bs{\Hat{v}}''$. 

The rotation matrix $\mathcal{R}$ is built combining only the identity and direct products of the vectors $\bs{V}$ and $\bs{V}''$ with themselves and with each other, hence, when applied to a vector $\bs{\Hat{b}}$ orthogonal to both, it returns the same vector. Thus, $\mathcal{R}$ must describe a rotation around the axis orthogonal to both $\bs{V}$ and $\bs{V}''$, unless the velocities are antiparallel, in which case the rotation is the identity. 

In other words, if a Lorentz transformation $L$ is the product of two boosts, the rotation $\mathcal{R}$ defined in Eq.~\eqref{eq:rot} is the minimal rotation 
taking the velocity $\bs{v}$ of the frame $\mathfrak{S}''$ relative to $\mathfrak{S}$ to the velocity $-\bs{v}''$, with 
$\bs{v}''$ the velocity of the frame $\mathfrak{S}$ relative to $\mathfrak{S}''$. 
Using the definition of the previous subsection \ref{subsub:minimalrot}, the rotation is $\mathcal{R}(\bs{\Hat{v}},-\bs{\Hat{v}}'')$.  
As we have established that it is impossible that $\bs{V}=\bs{V}''\neq \bs{0}$, $\mathcal{R}$ is never indeterminate of the form $\tfrac{0}{0}$. 

Since we have already deduced the axis (or better, the plane) of the rotation, in order to fully characterize the latter we need only provide the angle of rotation $\Theta$. 
This is most easily achieved noting that since $\bs{V}''=-\mathcal{R}\mycdot \bs{V}$, 
applying $\mathcal{R}$ in Eq.~\eqref{eq:rotb1b2} to $\bs{v}$, and imposing that it results in $-\bs{v}''$, one finds 
\beq
1-\frac{(\gamma_{u}+1)(\gamma_{u'}+1)(\gamma_{v}+1)}{(\gamma_u+\gamma_{u'}+\gamma_v+1)^2}(1-\bs{\Hat{v}}\mycdot\bs{\Hat{v}}'') = 0\ .
\eeq
Thus, the angle $\Theta$ of the rotation $\mathcal{R}$ and the supplementary angle $\Phi$ between the velocity 
$\bs{v}$ of the frame $\mathfrak{S}''$ relative to $\mathfrak{S}$ and the 
velocity $\bs{v}''$ of the frame $\mathfrak{S}$ relative to $\mathfrak{S}''$ satisfy
\beq
1+\cos\Theta=1-\cos\Phi = \frac{(\gamma_u+\gamma_{u'}+\gamma_v+1)^2}{(\gamma_{u}+1)(\gamma_{u'}+1)(\gamma_{v}+1)}\  .
\label{eq:Stapp}
\eeq

This formula lacks determining the sense of the rotation, which we provide below. 
If we choose the orientation of the rotation axis $\bs{\Hat{b}}$ to be such that $\bs{V},-\bs{V}''=\mathcal{R}\bs{V},\bs{\Hat{b}}$ form a right-handed triad, 
\bal
\sin\Theta=& -\frac{(\bs{V}\vecp \bs{V}'')\mycdot\bs{\Hat{b}}}{|\bs{V}''| |\bs{V}|}= 
-\frac{(\bs{V}\vecp \bs{V}'')\mycdot\bs{\Hat{b}}}{|\bs{V}|^2} =
\left[\frac{\gamma_{u'} +\gamma_v}{\gamma_{u}+1}\frac{\gamma_{u} 
+\gamma_v}{\gamma_{u'}+1}-1\right]\frac{(\bs{U}\vecp \bs{U}')\mycdot\bs{\Hat{b}}}{|\bs{V}|^2}
\nonumber 
\\
=&\ (\gamma_v-1)\frac{\gamma_u+\gamma_v+\gamma_{u'} +1}{(\gamma_{u}+1)(\gamma_{u'}+1)}\frac{(\bs{U}\vecp \bs{V})\mycdot\bs{\Hat{b}}}{{\gamma_v^2-1}}
\nonumber 
\\
=&\ \frac{\gamma_{u}+\gamma_{v} +\gamma_{u'}+1}{(\gamma_u+1)(\gamma_v+1)(\gamma_{u'}+1)}|\bs{U}||\bs{V}|\sin\phi .
\eal
This is Stapp's formula \cite{Stapp}. 
Here we used $\bs{U}\vecp\bs{U}'=\bs{U}\vecp\bs{V}$, and defined $\phi$ as the smaller angle between $\bs{U}$ and $\bs{V}$. It is a quantity measurable in 
$\mathfrak{S}$.  
We may express $\gamma_{u'}$ as a function of quantities observable in the same frame $\mathfrak{S}$, using 
\beq
\gamma_{u'}=\gamma_u\gamma_v-|\bs{U}||\bs{V}|\cos{\phi}.
\eeq
Finally, we find that the angle of the rotation satisfies \cite{Ritus1961}
\beq
\tan{\frac{\Theta}{2}}=\frac{\sin{\Theta}}{1+\cos{\Theta}}=\frac{|\bs{U}||\bs{V}|\sin\phi }{(1+\gamma_{u})(1+\gamma_v)-|\bs{U}||\bs{V}|\cos{\phi}}, 
\eeq
or better, in terms of hyperbolic functions, recalling that $|\bs{U}|=\sinh{\theta_u}$, $\gamma_u=\cosh{\theta_u}$, \etc, with $\theta$ the relativistic rapidity, 
\beq
\tan{\frac{\Theta}{2}}=\frac{\sin\phi }{\coth{\frac{\theta_u}{2}}\coth{\frac{\theta_v}{2}}-\cos{\phi}} .
\label{eq:mytheta}
\eeq
This equation allows to express the rotation in terms of quantities measured in $\mathfrak{S}$. 
It is a formula that may be useful whenever there is an actual reference frame $\mathfrak{S}'$, the velocity $\bs{u}$ of which $\mathfrak{S}$ will be able to measure, and it is known that the transformation from $\mathfrak{S}'$ and $\mathfrak{S}''$ is a pure boost.  


%
\subsubsection{Infinitesimal Thomas--Wigner rotation \label{app:inftw}}
A particularly interesting case, applicable to the problem under consideration of an accelerated reference frame, 
is when the second velocity $\bs{U}'\to \bs{\delta U}'$ is infinitesimal. 
Then the second boost takes the representation of a vector $\vec{C}'$ 
to the representation $\vec{C}''$ in an inertial frame infinitesimally close, the axes of which have been FW transported from the axes of 
$\mathfrak{S}'$, which is but the reference frame that, in the inertial frame $\mathfrak{S}$, is obtained from $\mathfrak{S}'$ by the velocity change 
$\bs{v}=\bs{u}+\bs{a}\delta t$ (in terms of Newtonian 3--velocities and acceleration). 
To first order in $\delta t$, 
\beq
\gamma_{\delta u'}\simeq 1, \qquad \gamma_v \simeq \gamma_u + \gamma_u^3 \bs{a}(t)\mycdot\bs{u}(t) \delta t,
\eeq
thus the relativistic velocities appearing in the combined Lorentz transformation satisfy 
\bsubs
\bal
\bs{V} =&\  \bs{U}+ \bs{\delta U}\simeq \bs{U}+
\bs{\delta U}'+\frac{\gamma_u^3 \bs{a}(t)\mycdot\bs{u}(t) \delta t}{\gamma_{u}+1}\bs{U},\\
\bs{V}'' \simeq&\ -\bs{U}-\gamma_{u}\bs{\delta U}' ,
\eal
\esubs
while the Newtonian velocities satisfy 
\bsubs
\bal
\bs{v} =\  \bs{u}+ \bs{\delta u} \simeq&\ 
 \bs{u}+
\bs{\delta u}'-\frac{\gamma_u^2 \bs{a}(t)\mycdot\bs{u}(t) \delta t}{\gamma_{u}+1}\bs{u},\\
\bs{v}'' \simeq&\ 
-\bs{u}-
\bs{\delta u}'-\bs{a}(t)\mycdot\bs{u}(t) \delta t \bs{u},
\eal
\esubs
and the angle of rotation in Eq.~\eqref{eq:mytheta} becomes 
\beq
\delta \Theta \simeq (\gamma_u-1)\delta \phi,
\label{eq:dtheta}
\eeq
with $\delta \phi$ the infinitesimal angle between $\bs{u}$ and $\bs{v}=\bs{u}+\bs{a}\delta t$. 
The axis of the rotation is orthogonal to $\bs{u}$ and $\bs{u}+\bs{a} \delta t$, hence it is in the direction $\bs{u}\vecp\bs{a}$. 
In terms of boosts and rotation matrices, we have: 
\beq
 B[\bs{\delta v'}]B[\bs{u}] = R[\bs{\delta \Theta}] B[\bs{u+\delta u}], 
\eeq
which we may rewrite as 
\beq
R^{-1}[\bs{\delta \Theta}] B[\bs{\delta v'}] =  B[\bs{u+\delta u}]B[-\bs{u}], 
\eeq
Thus, we have rederived Thomas's correct (up to a sign) statement \cite{Thomas1927} that a boost $B[-\bs{u}]$ followed by $B[\bs{u+\delta u}]$ is equivalent to an infinitesimal boost $B[\bs{\delta v'}]$
followed by an infinitesimal rotation $\bs{\delta \Theta}=-\frac{\gamma_u-1}{u^2} \bs{u}\vecp \bs{\delta u}$. 

%
\subsection{The commutation hexagon}
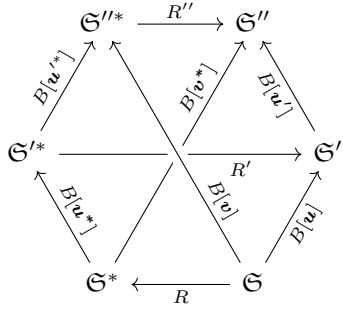
\begin{figure}[h!]
\centering
\begin{tikzcd}[column sep={1cm,between origins}, row sep={1.732050808cm,between origins}]
{}&\mathfrak{S}''^*\arrow[rr,"{R''}"]&{}&\mathfrak{S}''&{}\\
\mathfrak{S}'^*\ar[ur,"B{[}\bs{u}'^*{]}" sloped, pos=0.5]\arrow[rrrr, crossing over, "{R'}" near end, swap]&{}&{}&{}&\mathfrak{S}{'}\arrow[lu, "B{[}\bs{u}'{]}", sloped, pos=0.5, swap]\\
{}&\arrow[lu,"B{[}\bs{u^*}{]}" sloped, pos=0.5]\mathfrak{S}^*\arrow[rruu, crossing over, "B{[}\bs{v^*}{]}" sloped, pos=0.75]&{}&\arrow[ll,"R"]\mathfrak{S}\arrow[ru, "B{[}\bs{u}{]}", sloped, swap, pos=0.5]\arrow[lluu, crossing over, "B{[}\bs{v}{]}" sloped, pos=0.25]&{}
\end{tikzcd}
\caption[Diagrammatic representation of the commutation relation]{Diagrammatic representation of the commutation relations 
$B[\bs{u}'] B[\bs{u}] = R B[\bs{v}]= B[\bs{v}^*] R$.  
\label{fig:diagram}}
\end{figure}
In Fig.~\ref{fig:diagram}, we represent diagrammatically the results of the previous section by means of a hexagon, the vertices of which represent different reference frames. 
Two frames connected by a horizontal line are at rest relative to each other, differing only by a given rotation $R$. 
The rotation is the same for all frames, \ie\ it is represented by the same matrix. 
In the figure, 
$\bs{v}$ is the velocity of the double primed reference frames $\mathfrak{S}''$ and $\mathfrak{S}''^*$ relative to $\mathfrak{S}$, 
$\bs{u}'$ is the velocity of the double primed reference frames $\mathfrak{S}''$ and $\mathfrak{S}''^*$ relative to $\mathfrak{S}'$, and $\bs{u}$ is the velocity of the primed reference frames $\mathfrak{S}'$ and $\mathfrak{S}'^*$ relative to $\mathfrak{S}$. The three velocities are thus related through the relativistic composition law. Starred velocities are relative to the starred frame $\mathfrak{S}^*$. 
The common rotation 
$R=R'=R''=\left(\begin{smallmatrix} 1&\bs{0}^{\tr}\\ \bs{0}&\mathcal{R}\end{smallmatrix}\right)$ 
is a rotation of an angle 
$\Theta$ (defined univocally in \eqref{eq:mytheta}) in the common plane of $\bs{u}, \bs{v}, \bs{u}'$.  
The starred 3--vectors $\bs{u}^*, \bs{v}^*$ are obtained from the respective unstarred vectors through the rotation $\mathcal{R}$ while the primed and starred 
vector $\bs{u}'^*$ is obtained by the inverse rotation $\mathcal{R}^{-1}\bs{u}'$. 
Thus the hexagon is determined by specifying any two velocities, \eg\ $\bs{u}$ and $\bs{u}'$.
%

\section{\label{sec:diagrams}The meaning of Thomas--Wigner rotation revealed through commutative diagrams}
The pitfalls that lead to the error in Thomas precession could be avoided by using commutative diagrams for Lorentz transformations as a visual aid. 
It is well known that a Lorentz transformation from an inertial frame $\mathfrak{S}$ to an inertial frame $\mathfrak{S}'$ moving with velocity $\bs{v}$ relative to 
$\mathfrak{S}$ can be decomposed as the product of a boost and a rotation in two ways: $L=R B[\bs{v}]$ and $L=B[-\bs{v}'] R$, with $\bs{v}'$ the velocity of $\mathfrak{S}$ relative to $\mathfrak{S}'$. 
We apply these decompositions to the series of Lorentz transformations from the lab frame to the comoving FW transported frames $\mathfrak{S}'_t$ 
the origins of which coincide with the position of an accelerated particle. Some care must be taken because dynamical quantities measured in the accelerated 
frame do not coincide with corresponding quantities measured in the instantaneous rest frame $\mathfrak{S}'_t$, since they involve incremental ratios 
of quantities measured in two different inertial frames $\mathfrak{S}'_t$ and $\mathfrak{S}'_{t+\delta t}$. 

In the diagrams, boosts are indicated by dashed straight lines, rotations by dotted curved lines, and 
general Lorentz transformations by full curved lines. Two additional frames appear: a comoving frame, the axes of which rotate relative to the comoving FW 
frame, which we will call the comoving boosted frame $\mathfrak{S}'^*$; and a frame with the origin at rest in the lab, but rotating axes, which we will refer to as the rotating lab frame $\mathfrak{S}^*$. 
In the literature, the distinction between the FW frame and the rotating frame has been seldom recognized, with a few exceptions, as for instance the excellent book by Garg \cite{Garg}. 

\begin{figure}[h!]
\centering
\begin{tikzcd}[column sep={2cm,between origins}, row sep={1.732050808cm,between origins}]
\mathfrak{S}'^*_{t_0+\delta t}\arrow[r, dotted, bend right=20, swap, "\delta R'_{t'_0}"]\arrow[rr, dotted, bend left = 30, "R'_{t'_0+\delta t'}"]&\mathfrak{S}''_{t_0+\delta t}\arrow[r, dotted, bend right=20, swap, "R'_{t'_0}"]&[+15ex]\mathfrak{S}'_{t_0+\delta t}\\
&\mathfrak{S}'^*_{t_0}=\mathfrak{S}''_{t_0}\arrow[u,dashed, "B{[}\bs{\delta v}'^*{]}" {sloped,rotate=180,yshift=.5ex,pos=0.5}]\arrow[r, dotted, bend right=20, "R'_{t'_0}"]&\mathfrak{S}'_{t_0}\arrow[u, dashed, "B{[}\bs{\delta v}'{]}" {xshift=-1.5ex, rotate = -90, pos=0}]\\[10ex] 
&\arrow[luu, dashed, "B{[}\bs{v}_{t_0+\delta t}{]}"  rotate=0, swap,sloped, pos=0.5]\mathfrak{S}\arrow[u,dashed, "B{[}\bs{v}_{t_0}{]}" {sloped,rotate=180,yshift=.5ex,pos=0.5}]\arrow[ru, "L_{t_0}", bend right = 30]\arrow[ruu, bend right = 70, swap, "L_{t_0+\delta t}"]&
\end{tikzcd}
\caption{Commutative diagram for the polar decomposition of Lorentz transformations $L=R B[\bs{v}]$. 
 We considered a generic time $t=t_0$ in the lab frame $\mathfrak{S}$, corresponding to a time $t'_0$ in any comoving frame, and an infinitesimal time interval $\delta t$, which 
is $\delta t'$ in any comoving frame. The primed and starred sequence of frames $\mathfrak{S}'^*_t$ is defined by applying pure Lorentz boosts to the lab frame. The primed sequence of frames $\mathfrak{S}'_t$ is obtained by FW transport. The double primed sequence of frames 
$\mathfrak{S}''_{t}$ is an auxiliary FW frame, the axes of which 
are rotated relative to $\mathfrak{S}'_{t}$ by the constant rotation $\mathcal{R}^{-1}_{t'_0}$.
\label{fig:cd}}
\end{figure}
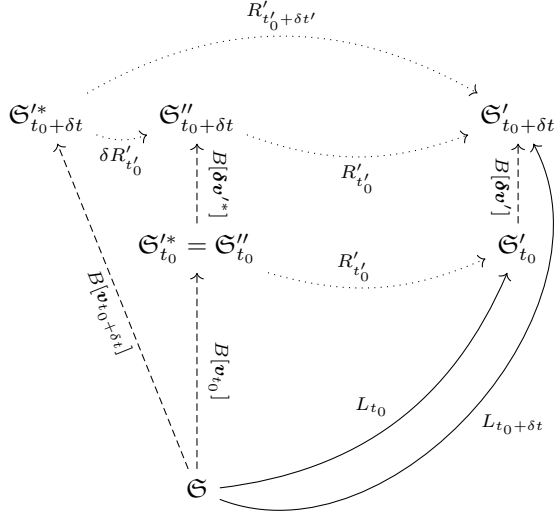

With reference to Fig.~\ref{fig:cd},
let $\mathfrak{S}'_t$ a FW frame comoving with the accelerated particle. Let $L_t$ the Lorentz transformations from the lab frame to the comoving 
FW frame at lab time $t$. Let $B[\bs{v}]$ the Lorentz boost from the lab frame to a frame centered at the particle and moving with velocity $\bs{v}$ relative to it, and let 
$\mathfrak{S}'^*_t$ the comoving frame defined by the boosts $B[\bs{v}_t]$ as the velocity of the particle varies in time. 
As the FW frame and the comoving boosted frame are at rest relative to each other, they are connected by a time--dependent rotation $R'_{t'}$, 
the finite Thomas--Wigner rotation, 
with $t'$ their common proper time, a known function of $t$, so that $L_t = R'_{t'} B[\bs{v}_t]$. 
While there is no general analytic expression for $R'_{t'}$, its infinitesimal value is, as found by Thomas \cite{Thomas1927} (see the former section \ref{app:inftw} for 
a derivation with the correct sign)
\beq
\delta R'_{t'}=B[\bs{\delta v}'^*]B[\bs{v}_{t_0}]B[-\bs{v}_{t_0+\delta t}]=\id_4+\frac{\gamma_t}{\gamma_t+1} (\bs{v}_t\vecp \bs{\delta v}'^*)\mycdot\bs{J},
\label{eq:twrot}
\eeq
 with $\bs{J}=\left(\begin{smallmatrix}1&\bs{0}^\tr\\ \bs{0}&\bs{\mathcal{J}}\end{smallmatrix}\right)$ the $4\times 4$ generators of spatial rotations and $\bs{\delta v}'^*=\bs{a}'^*_{t'}\delta t'$ the infinitesimal velocity increment measured in the boosted frame, $\bs{a}'^*$ being the proper acceleration measured in the comoving boosted frame. 
%

From the commutative diagram Fig.~\ref{fig:cd}, we infer that the 
rotation satisfies 
\beq
R'_{t'+\delta t'}=R'_{t'} \delta R'_{t'}  \ , 
\label{eq:diffrot}
\eeq 
from which it follows immediately that the instantaneous angular velocity of the axes of $\mathfrak{S}'^*_{t}$ relative to the axes of $\mathfrak{S}'_t$ is 
\beq
\bs{\omega}'_{t'}=\frac{\gamma_t}{\gamma_t+1} 
\mathcal{R}_{t'}(\bs{v}_t\vecp \frac{\bs{\delta v}'^*}{\delta t'})=
-\frac{\gamma_{t}}{\gamma_{t}+1} \bs{v}'_{t}\vecp\bs{a}'_{t},
\label{eq:spec1}
\eeq
where we used the fact that $\bs{v}_t=-\bs{v}'^*_t$, the velocity of the frame $\mathfrak{S}$ relative to 
$\mathfrak{S}'^*$ is the opposite of $\bs{v}_t$, since the frames are connected by a boost, and 
the fact that applying $\mathcal{R}_{t'}$ to a vector of the boosted 
comoving frame $\mathfrak{S}'^*$ yields the corresponding vector in the comoving FW frame $\mathfrak{S}'$. 
The acceleration $\bs{a}'_{t}$ is thus the proper acceleration of the FW frame $\mathfrak{S}'$, 
while $\bs{v}'_{t}$ is the velocity of $\mathfrak{S}$ relative to 
$\mathfrak{S}'$, in the sense that it is the velocity of 
the point of $\mathfrak{S}$ passing through the origin of $\mathfrak{S}'$ at time $t'$. Points at rest in an inertial frame appear to move, in an accelerated frame, with a position dependent velocity, similar to Hubble's law, in addition to the Coriolis velocity \cite{DiLorenzoprep}.\label{fn:hubble} Thus, when stating the velocity of an inertial frame relative to an accelerated frame, one must specify which point of the inertial frame is being considered. 
The rotation of the axes of $\mathfrak{S}'_{t}$ relative to the axes of $\mathfrak{S}'^*_t$ is, instead 
\beq
\bs{\omega}'^*_{t'}=
\frac{\gamma_{t}}{\gamma_{t}+1} \bs{v}'^*_{t}\vecp\bs{a}'^*_{t}.
\label{eq:spec1bapp}
\eeq
In terms of quantities measured in the lab frame
\beq
\bs{\omega}'^*_{t'}=
-\frac{\gamma^3_{t}}{\gamma_{t}+1} \bs{v}_{t}\vecp\bs{a}_{t}.
\label{eq:spec1c}
\eeq
Equation \eqref{eq:spec1c} is the source of a common misunderstanding on the Thomas--Wigner rotation: 
the rate on the lhs is observed in the boosted comoving frame of the particle, but the rhs is expressed as 
a function of quantities measured in the lab frame. 
This means merely that an inertial observer can predict what the accelerated observer will be measuring. 
However, a common mistake is to consider the lhs as if it was a quantity observed in the lab frame, and hence 
consider the rate in the lab frame, $\bs{\omega}'^*_{t}=\bs{\delta \Phi}/dt = 
-\frac{\gamma^2_{t}}{\gamma_{t}+1} \bs{v}_{t}\vecp\bs{a}_{t}$. This is the equation for Thomas precession, as most commonly 
derived.

\begin{figure}[h!]
\centering
\begin{tikzcd}[column sep={2cm,between origins}, row sep={1.732050808cm,between origins}]
&[+15ex]\mathfrak{S}'_{t_0+\delta t}&\\
&[+15ex]\mathfrak{S}'_{t_0}\arrow[u,dashed, "B{[}\bs{\delta v}'{]}" {swap,sloped,rotate=180,yshift=-1.ex,pos=0.5}]&
\\[10ex] 
\mathfrak{S}\arrow[rr, dotted, bend right=20, "R_{t_0+\delta t}"]\arrow[r, dotted, bend left=20, "R_{t_0}"]\arrow[ru, "L_{t_0}", bend left = 30]\arrow[ruu, bend left = 50, swap, "L_{t_0+\delta t}"]
&[+15ex]\mathfrak{S}^*_{t_0}\arrow[u,dashed, "B{[}\mathcal{R}_{t_0}\bs{v}_{t_0}{]}" {swap,sloped,rotate=180,yshift=-1.ex,pos=0.5}]
\arrow[r, dotted, bend left=20, "\delta R_{t_0}", pos=0.6]&\mathfrak{S}^*_{t_0+\delta t}
\arrow[uul,dashed, "B{[}\mathcal{R}_{t_0+\delta t}\bs{v}_{t_0+\delta t}{]}" {sloped, pos=0.5}]
\end{tikzcd}
\caption{Commutative diagram for the alternative polar decomposition of Lorentz transformations $L = B[\mathcal{R}\bs{v}] R$. 
Here $\mathfrak{S}^*_t$ is a rotating frame with the origin at rest in the lab, defined as the frame obtained by applying the boosts $B[\bs{v}'_{t'}]$ to the comoving FW frame, with $\bs{v}'_{t'}$ being the velocity of the lab frame relative to the particle at time $t'=\tau(t)$. 
\label{fig:cd2}}
\end{figure}
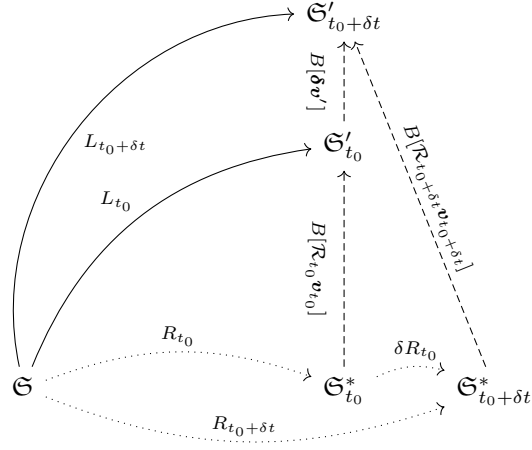

We may proceed analogously for the alternative decomposition of Lorentz transformations, $L=BR$. 
In the commutative diagram \ref{fig:cd2}, while the rotation is represented by the same matrix $R$ as in the previous figure, 
it is a physically distinct operation, as it describes  
the rotation of the axes of $\mathfrak{S}^*_t$ relative to the axes of the lab frame $\mathfrak{S}$. Hence we are denoting it as $R_t$, with the time $t$ measured in the lab frame and the rotating frame.  
From the commutative diagram Fig.~\ref{fig:cd2}, we infer that the rotation satisfies 
\beq
R_{t+\delta t}=\delta R_{t} R_{t} \ .
\label{eq:diffrot2}
\eeq 
The matrices $R_t$ and $R'_{t'}$ coincide, for any $t'=\tau(t)$ simultaneous with the lab time $t$  (according to the simultaneity of the lab frame: we recall that simultaneity in special relativity is not a reciprocal property). 
However, in this second decomposition the infinitesimal Thomas--Wigner rotation $\delta R_t$  does not coincide with 
$\delta R'_{t'}$ in Eq.~\eqref{eq:twrot}, but, as is clear comparing diagrams \ref{fig:cd} and \ref{fig:cd2}, it is 
\beq
\delta R_{t}=B[-\bs{v}^*_{t_0+\delta t}]B[\bs{\delta v}']B[\bs{v}^*_{t_0}]=\id_4+\frac{\gamma_t}{\gamma_t+1} [(\mathcal{R}_t\bs{v}_t)\vecp \bs{\delta v}']\mycdot\bs{J},
\label{eq:twrot2}
\eeq
Indeed, in order for the matrices $R_t$ and $R'_{t'}$ to coincide identically in time, since they obey distinct differential equations 
$R'_{t'} \delta R' =R'_{t'+\delta t'}$ and $\delta R R_{t}  =R_{t+\delta t}$, the increments $\delta R$ and $\delta R'$ must be conjugated through $R$, 
$\delta R' = R^{-1} \delta R R$. 

The axes of $\mathfrak{S}^*_{t}$ relative to the axes of $\mathfrak{S}$ transform with the inverse matrix, 
$\mathcal{R}_t^{-1}$. 
It follows immediately that the instantaneous angular velocity of the axes of $\mathfrak{S}^*_{t}$ relative to the axes of $\mathfrak{S}$ is 
\beq
\bs{\omega}_{t}=-\frac{\gamma_t}{\gamma_t+1} \bs{v}_t \vecp \mathcal{R}^{-1}_t\frac{\bs{\delta v}'}{\delta t}=
-\frac{\gamma_t^2}{\gamma_t+1} \bs{v}_t\vecp \bs{a}_t\ .
\label{eq:spec2}
\eeq
Here, we used $\gamma_t\mathcal{R}^{-1}_t\frac{\bs{\delta v}'}{\delta t}=\mathcal{R}^{-1}_t\frac{\bs{\delta v}'}{\delta t'}=\bs{a}'^*_{t'}=\gamma^ 2_t \bs{a}_t + \frac{\gamma_t^4}{\gamma_t+1} \bs{v}_t\mycdot\bs{a}_t \bs{v}_t$, 
since $\bs{a}'^*_{t'}$ is the proper acceleration in the comoving boosted frame. 

Finally, we have a quantity which is measured in the lab frame, given by Eq.~\eqref{eq:spec2}: it is the angular velocity, relative to the lab frame, of a purely rotating frame, whose origin is at rest in the lab. 
The rotation \eqref{eq:spec2} guarantees that the Lorentz transformation from the rotating frame to the FW accelerated frame is a Lorentz boost.

In summary, the Thomas--Wigner rotation 
describes the instantaneous rotation of a frame relative to another.
It does not describe the rotation of a vector transported by an accelerated particle as seen in the lab frame, 
nor an additional Coriolis rotation seen by the accelerated particle. 
Rather, it describes the Coriolis rotation observed either in a rotating frame with the origin at rest relative to the lab frame, or in a frame comoving with the particle but the axes 
of which are not FW transported. 
Therefore, the Thomas--Wigner rotation rate while strictly related to the Thomas precession --- defined as the rotation observed in the lab frame of the spin of  a particle due 
only to relativistic kinematics effects --- does not coincide with it, 
and thus can not possibly account for the energy levels of an atom at rest in the lab frame. 

 %
 \section{The interplay of Thomas--Wigner rotation and Lorentz contraction in the spin dynamics\label{app:interplay}} 
\begin{figure}[h]
     \centering
     \begin{subfigure}[b]{0.3\textwidth}
         \centering
         \includegraphics[width=\textwidth]{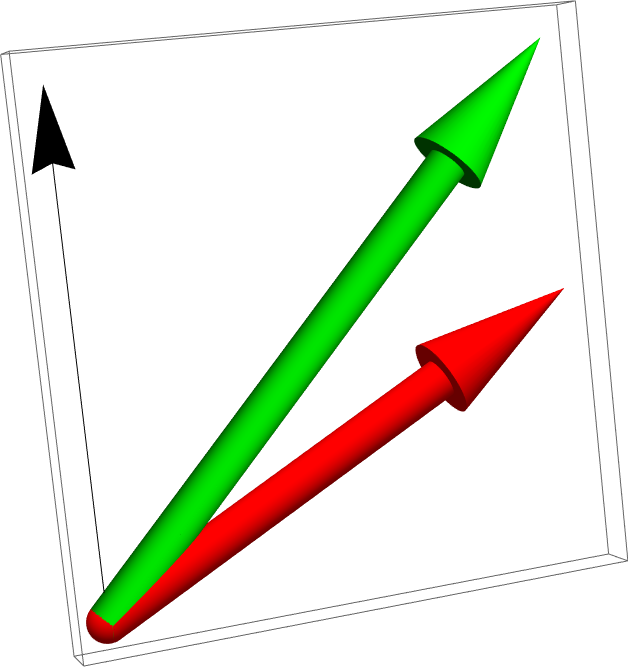}
\end{subfigure}
    \begin{subfigure}[b]{0.3\textwidth}
         \centering
         \includegraphics[width=0.18\textwidth]{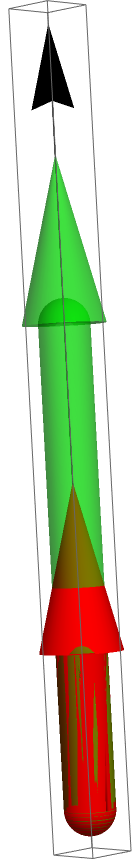}
\end{subfigure}
    \begin{subfigure}[b]{0.3\textwidth}
         \centering
         \includegraphics[width=\textwidth]{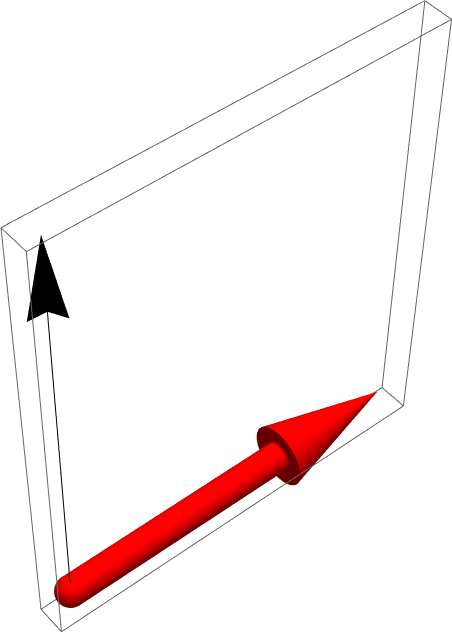}
\end{subfigure}
\caption{\label{fig:lorentzrot} A 3D vector which is represented as the green arrow in a reference frame, is represented by the red arrow in a new reference frame boosted with velocity $\bs{v}$ (black arrow) with respect to the former frame. 
In general, the red vector will be rotated relative to the green vector, because of Lorentz contraction. 
Here, we chose a Lorentz factor $\gamma=2$ for illustrative purposes.}
\end{figure}

First of all, we establish a simple fact: pure Lorentz boosts imply a nonlinear rotation. 
Indeed, let us consider first a vector $\bs{P}$ associated to a particle initially at rest in the lab frame. 
If the particle is boosted by a velocity $\bs{\Delta v}$, the vector 
$\bs{P}$ will contract in the direction of the velocity, as illustrated in 
the figure \ref{fig:lorentzrot}. Therefore, the vector $\bs{P}$ will rotate by an angle $\Delta \theta$, depending on its orientation relative to the velocity boost. 

\begin{figure}
\includegraphics[width=\textwidth]{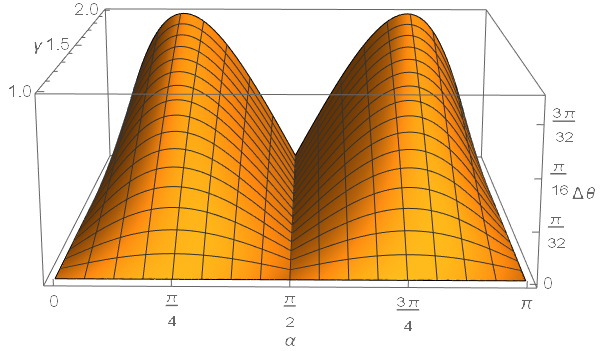}
\caption{\label{fig:lorentzrotangle} The angle of rotation $\Delta\theta$ of a vector $\bs{P}$ due to Lorentz contraction as a function of the angle $\alpha$ that the vector makes with the velocity of the boost, for several values of $\gamma$.}
\end{figure}

The angle $\Delta \theta$ is zero when $\bs{P}$ is normal, parallel, or antiparallel  to the direction of the boost, and it reaches a maximum when the angle $\alpha$ between $\bs{P}$ and $\bs{\Delta v}$ satisfies 
$\cos(2\alpha)=\frac{\gamma_{\Delta v}-1}{\gamma_{\Delta v}+1}$, which yields $\alpha\to 0,\pi$ in the ultra--relativistic limit, and $\alpha\simeq \pi/4$ in the non--relativistic limit (Fig. \ref{fig:lorentzrotangle}). 

This rotation combines with the rotation of the vector due to external influences, and also with the Thomas--Wigner rotation. 
Precisely, we have shown that the Lorentz transformation from the lab frame to the FW rest frame of the particle is $L_\mathrm{FW}(t)=R_\mathrm{TW}(t)B[\bs{v}_t]$. 
Hence, since in the rest frame the spin is represented as the 4--vector: 
$\{0,\bs{S}'\}$, the corresponding representation in the lab frame is 
\beq
\bp
S^0(t)\\
\bs{S}_t\ep
=B[-\bs{v}_t]R^\tr_\mathrm{TW}(t)
\bp
0\\
\bs{S}'_t\ep ,
\eeq
where we used the fact that the inverse of a Lorentz boost $B[\bs{v}]$ is 
$B^{-1}[\bs{v}]=B[-\bs{v}]$, and the inverse of a rotation is 
$R^{-1}=R^\tr$. 

The relevant spatial part is 
\beq
\bs{S}_t = \mathcal{R}^\tr_\mathrm{TW}(t)\bs{S}'_t+(\gamma_t-1)\bs{\Hat{v}}_t\bs{\Hat{v}}_t^\tr  \mathcal{R}^\tr_\mathrm{TW}(t)\bs{S}'_t .
\label{eq:3s}
\eeq
Here, we used Eqs.~\eqref{eq:br}. 
The second term on the rhs describes the dilation of the vector $\bs{S}$ along the velocity axis. 

For the simultaneity--corrected vector $\bs{T}_t$, instead 
\beq
\bs{T}_t = \mathcal{R}^\tr_\mathrm{TW}(t)\bs{S}'_t+(\gamma^{-1}_t-1)\bs{\Hat{v}}_t\bs{\Hat{v}}_t^\tr  \mathcal{R}^\tr_\mathrm{TW}(t)\bs{S}'_t .
\label{eq:3t}
\eeq
The second term on the rhs describes the contraction of the vector $\bs{T}$ along the velocity axis. 

Let us assume temporarily that we could neglect the effects of Lorentz contraction/dilation. 
Then 
\beq
\bs{S}_t \simeq\bs{T}_t \simeq \mathcal{R}^\tr_\mathrm{TW}(t)\bs{S}'_t .
\eeq
In the comoving FW frame, the spin $\bs{S}'$ makes a Larmor precession due to the electromagnetic field acting on the particle therein, 
\beq
\frac{d}{dt'}\bs{S}'_{t'} = \bs{\omega}'_\mathrm{L}(t)\times \bs{S}'_{t'}=\Omega'_\mathrm{L}(t) \bs{S}'_{t'} ,
\eeq
where $\Omega'_\mathrm{L}$ is the antisymmetric matrix associated to the vector $\bs{\omega}'_\mathrm{L}$. 
In the lab frame, therefore, neglecting consistently, within the approximation of the present paragraph, the difference between the factor $\gamma_t=\frac{dt'}{dt}$ and 1, 
\bal
\frac{d}{dt}\bs{S}_t  &\simeq 
\left[ \mathcal{R}^\tr_\mathrm{TW}(t)\Omega'_\mathrm{L} \mathcal{R}_\mathrm{TW}(t)
+\Omega_\mathrm{TW}(t) \right]\mathcal{R}^\tr_\mathrm{TW}(t)\bs{S}'_t
\nonumber\\
&= \left[\Omega_\mathrm{L}(t)+\Omega_\mathrm{TW}(t)\right]\bs{S}_t = 
\left[\bs{\omega}_\mathrm{L}(t)+\bs{\omega}_\mathrm{TW}(t)\right]\times\bs{S}_t.
\eal
Here, we introduced the Larmor frequency in the lab frame, 
$\Omega_\mathrm{L}(t)=\mathcal{R}^\tr_\mathrm{TW}(t)\Omega'_\mathrm{L} \mathcal{R}_\mathrm{TW}(t)$, 
$\bs{\omega}_\mathrm{L}(t)=\mathcal{R}^\tr_\mathrm{TW}(t) \bs{\omega}'_\mathrm{L}(t)$ and the Thomas--Wigner rotation frequency, defined by $\frac{d}{dt}\mathcal{R}^\tr_\mathrm{TW}(t)=\Omega_\mathrm{TW}(t)\mathcal{R}^\tr_\mathrm{TW}(t)$. From Eq.~\eqref{eq:dtheta}, with $\bs{v}_t\to\bs{u}$, we find 
\bal
\bs{\omega}_\mathrm{TW}(t):=-\frac{{\delta\Theta}}{\delta t}\frac{\bs{v}_t\times \bs{a}_t}{|\bs{v}_t\times \bs{a}_t|}&= 
-(\gamma_t-1)\frac{\bs{v}_t\times(\bs{v}_t+\bs{a}_t\delta t)}{|\bs{v}_t\times(\bs{v}_t+\bs{a}_t\delta t)|} \frac{\delta\phi}{\delta t}
\nonumber \\
&\simeq
-(\gamma_t-1)\frac{\bs{v}_t\times\bs{a}_t\delta t}{\bs{v}^2_t\sin(\delta\phi)} \frac{\delta\phi}{\delta t}
\simeq 
-(\gamma_t-1)\frac{\bs{v}_t\times\bs{a}_t}{\bs{v}^2_t} .
\eal
Due to the presence of $\gamma_t-1$, the additional Thomas--Wigner rotation, however, is of the same order of the contributions due to the Lorentz contraction/dilation which were neglected. Thus, the approximation is not consistent. Furthermore, we wish to know the dynamics of the spin also in the relativistic regime. 

Therefore, we need the full Eq.~\eqref{eq:3s} or Eq.~\eqref{eq:3t}, which do 
not conserve the norm of the spin 3--vector. In these equations, the effects 
of the Thomas--Wigner rotation $\mathcal{R}_\mathrm{TW}$ and of the Lorentz contraction/dilation $(\gamma_t^{\pm 1}-1)\bs{\Hat{v}}_t\bs{\Hat{v}}_t^\tr$ 
get twined. 

Indeed, let $\mathcal{L}^+_t = \id_3+(\gamma_t-1)\bs{\Hat{v}}_t\bs{\Hat{v}}_t^\tr$ the $3\times 3$ matrix describing Lorentz dilation, 
$\mathcal{L}^{-}_t = \id_3+(\gamma^{-1}_t-1)\bs{\Hat{v}}_t\bs{\Hat{v}}_t^\tr$ the $3\times 3$ matrix, inverse of $\mathcal{L}^+$, describing Lorentz contraction, 
and $\bs{P}^+=\bs{S}$, $\bs{P}^-=\bs{T}$. 
Upon differentiation of Eq.~\eqref{eq:3s} and Eq.~\eqref{eq:3t}, there are terms which involve $\mathcal{L}^\pm$, $\mathcal{R}_\mathrm{TW}$, and their derivatives: 
\beq
\frac{d}{dt}\bs{P}^\pm_t =
 [\frac{d}{dt}\mathcal{L}^\pm_t] {\mathcal{L}^\mp_t} \bs{P}^\pm_t +
\mathcal{L}^\pm_t \Omega_\mathrm{TW}(t){\mathcal{L}^\mp_t}
\bs{P}^\pm_t +\Omega^\pm_\mathrm{L}(t) \bs{P}^\pm_t,
\eeq
with the Larmor matrix in the lab frame $\Omega^\pm_\mathrm{L}=\gamma \mathcal{L}^\pm \mathcal{R}^\tr_\mathrm{TW}\Omega'_\mathrm{L}\mathcal{R}_\mathrm{TW} \mathcal{L}^\mp$, no longer a skew--symmetric matrix as $\Omega'_\mathrm{L}$. 

Thus, one cannot say ``this term is due to the Lorentz contraction, this other term is due to the Thomas--Wigner rotation.''
In order to extract a rotation, one should 
use the technique illustrated in the Methods section of the main manuscript.
%
\bibliographystyle{naturemag}

\begin{thebibliography}{99}
\expandafter\ifx\csname url\endcsname\relax
  \def\url#1{\texttt{#1}}\fi
\expandafter\ifx\csname urlprefix\endcsname\relax\def\urlprefix{URL }\fi
\expandafter\ifx\csname doiprefix\endcsname\relax\def\doiprefix{DOI: }\fi
\providecommand{\bibinfo}[2]{#2}
\providecommand{\eprint}[2][]{\url{#2}}

\bibitem{Thomas1926}
\bibinfo{author}{Thomas, L.~H.}
\newblock \bibinfo{journal}{\bibinfo{title}{The motion of the spinning
  electron}}.
\newblock {\emph{{Nature}}} \textbf{\bibinfo{volume}{117}},
  \bibinfo{pages}{514} (\bibinfo{year}{1926}).

\bibitem{Thomas1927}
\bibinfo{author}{Thomas, L.~H.}
\newblock \bibinfo{journal}{\bibinfo{title}{I. the kinematics of an electron
  with an axis}}.
\newblock {\emph{{Philos.~Mag.~Ser.~7}}}
  \textbf{\bibinfo{volume}{3}}, \bibinfo{pages}{1} (\bibinfo{year}{1927}).

\bibitem{Jackson}
\bibinfo{author}{Jackson, J.~D.}
\newblock \emph{\bibinfo{title}{Classical Electrodynamics}}
  (\bibinfo{publisher}{{John Wiley and Sons}}, \bibinfo{year}{1999}),
  \bibinfo{edition}{3rd} edn.

\bibitem{Malykin}
\bibinfo{author}{Malykin, G.~B.}
\newblock \bibinfo{journal}{\bibinfo{title}{Thomas precession: correct and
  incorrect solutions}}.
\newblock {\emph{{Phys.-Usp.}}} \textbf{\bibinfo{volume}{49}},
  \bibinfo{pages}{837} (\bibinfo{year}{2006}).

\bibitem{Ritus}
\bibinfo{author}{Ritus, V.~I.}
\newblock \bibinfo{journal}{\bibinfo{title}{On the difference between
  {Wigner's} and {M{\o}ller's} approaches to the description of {Thomas}
  precession}}.
\newblock {\emph{{Phys.-Usp.}}} \textbf{\bibinfo{volume}{50}},
  \bibinfo{pages}{95} (\bibinfo{year}{2007}).

\bibitem{Farago}
\bibinfo{author}{Farago, P.~S.}
\newblock \bibinfo{journal}{\bibinfo{title}{The polarization of electron beams
  and the measurement of the g-factor anomaly of free electrons}}.
\newblock {\emph{{Adv. Electron Electron. Phys.}}}
  \textbf{\bibinfo{volume}{21}}, \bibinfo{pages}{1--66} (\bibinfo{year}{1966}).

\bibitem{Garg}
\bibinfo{author}{Garg, A.}
\newblock \emph{\bibinfo{title}{Classical Electromagnetism in a Nutshell}}
  (\bibinfo{publisher}{Princeton University Press}, \bibinfo{year}{2012}).

\bibitem{Stepanov2012}
\bibinfo{author}{Stepanov, S.~S.}
\newblock \bibinfo{journal}{\bibinfo{title}{Thomas precession for spin and for
  a rod}}.
\newblock {\emph{{Phys. Part. Nucl.}}}
  \textbf{\bibinfo{volume}{43}}, \bibinfo{pages}{128--145}
  (\bibinfo{year}{2012}).
\newblock \bibinfo{note}{[Original Russian: Fiz. Elementar. Chastits i Atom.
  Yadra \textbf{43}, 245--282 (2012)]}.

\bibitem{BMT}
\bibinfo{author}{Bargmann, V.}, \bibinfo{author}{Michel, L.} \&
  \bibinfo{author}{Telegdi, V.~L.}
\newblock \bibinfo{journal}{\bibinfo{title}{Precession of the polarization of
  the particles moving in a homogeneous electromagnetic field}}.
\newblock {\emph{{Phys.~Rev.~Lett.}}}
  \textbf{\bibinfo{volume}{2}}, \bibinfo{pages}{435} (\bibinfo{year}{1959}).

\bibitem{Kholmetskii2020}
\bibinfo{author}{Kholmetskii, A.~L.} \& \bibinfo{author}{Yarman, T.}
\newblock \bibinfo{journal}{\bibinfo{title}{The relativistic mechanism of the
  {Thomas--Wigner} rotation and {Thomas} precession}}.
\newblock {\emph{{Eur.~J.~Phys.}}} \textbf{\bibinfo{volume}{41}},
  \bibinfo{pages}{055601} (\bibinfo{year}{2020}).

\bibitem{Borel}
\bibinfo{author}{\'{E}mile Borel}.
\newblock \bibinfo{journal}{\bibinfo{title}{La th\'{e}orie de la relativit\'{e}
  et la cin\'{e}matique}}.
\newblock {\emph{{C.~r.~hebd.~Seanc.~Acad.~Sci.}}}
  \textbf{\bibinfo{volume}{156}}, \bibinfo{pages}{215} (\bibinfo{year}{1913}).

\bibitem{Borel2}
\bibinfo{author}{\'{E}mile Borel}.
\newblock \emph{\bibinfo{title}{Introduction geom\'{e}trique a quelques
  th\'{e}ories physiques}} (\bibinfo{publisher}{Gauthiers--Villar},
  \bibinfo{year}{1914}).

\bibitem{Minkowski}
\bibinfo{author}{Minkowski, H.}
\newblock \bibinfo{journal}{\bibinfo{title}{Raum und {Zeit}}}.
\newblock {\emph{{Physik. Z.}}} \textbf{\bibinfo{volume}{10}},
  \bibinfo{pages}{104} (\bibinfo{year}{1909}).
\newblock \bibinfo{note}{[English transl. \emph{Space and Time}, in {Spacetime:
  Minkowski's Papers on Spacetime Physics}, (Minkowski Institute Press,
  2021)]}.

\bibitem{MTW}
\bibinfo{author}{Misner, C.~W.}, \bibinfo{author}{Thorne, K.~S.} \&
  \bibinfo{author}{Wheeler, J.~A.}
\newblock \emph{\bibinfo{title}{Gravitation}} (\bibinfo{publisher}{Princeton
  University Press}, \bibinfo{year}{2017}).

\bibitem{Fermi}
\bibinfo{author}{Fermi, E.}
\newblock \bibinfo{journal}{\bibinfo{title}{Sopra i fenomeni che avvengono in
  vicinanza di una linea oraria}}.
\newblock {\emph{{Rend.~Lincei}}} \textbf{\bibinfo{volume}{31}},
  \bibinfo{pages}{21} (\bibinfo{year}{1922}).
\newblock \bibinfo{note}{\emph{ibid.}, 51; \emph{ibid.} 101; {collected and
  reprinted in} \emph{Enrico Fermi, Collected papers}, vol.~I, University of
  Chicago Press (1962)}.

\bibitem{Walker}
\bibinfo{author}{Walker, A.~G.}
\newblock \bibinfo{journal}{\bibinfo{title}{Relative co-ordinates}}.
\newblock {\emph{{Proc.~R.~Soc.~Edinb.}}}
  \textbf{\bibinfo{volume}{52}}, \bibinfo{pages}{345} (\bibinfo{year}{1933}).

\bibitem{muon}
\bibinfo{author}{Bennett, G.~W.} \emph{et~al.}
\newblock \bibinfo{journal}{\bibinfo{title}{Final report of the {E821} muon
  anomalous magnetic moment measurement at {BNL}}}.
\newblock {\emph{{Phys.~Rev.~D}}} \textbf{\bibinfo{volume}{73}},
  \bibinfo{pages}{072003} (\bibinfo{year}{2006}).

\bibitem{electroncyclotron}
\bibinfo{author}{Odom, B.}, \bibinfo{author}{Hanneke, D.},
  \bibinfo{author}{{D’Urso}, B.} \& \bibinfo{author}{Gabrielse, G.}
\newblock \bibinfo{journal}{\bibinfo{title}{New measurement of the electron
  magnetic moment using a one-electron quantum cyclotron}}.
\newblock {\emph{{Phys.~Rev.~Lett.}}}
  \textbf{\bibinfo{volume}{97}}, \bibinfo{pages}{030801}
  (\bibinfo{year}{2006}).

\bibitem{muon2}
\bibinfo{author}{Abi, B.} \emph{et~al.}
\newblock \bibinfo{journal}{\bibinfo{title}{Measurement of the positive muon
  anomalous magnetic moment to 0.46 ppm}}.
\newblock {\emph{{Phys. Rev. Lett.}}}
  \textbf{\bibinfo{volume}{126}}, \bibinfo{pages}{141801},
  \doiprefix\url{10.1103/PhysRevLett.126.141801} (\bibinfo{year}{2021}).

\bibitem{Einstein1907}
\bibinfo{author}{Einstein, A.}
\newblock \bibinfo{journal}{\bibinfo{title}{\"{U}ber das relativit\"{
  a}tsprinzip und die aus demselben gezogene folgerungen}}.
\newblock {\emph{{Jahrbuch der Radioaktivit\"{a}t und
  Elektronik}}} \textbf{\bibinfo{volume}{4}}, \bibinfo{pages}{411}
  (\bibinfo{year}{1907}).
\newblock \bibinfo{note}{[English transl. \emph{On the relativity principle and
  the conclusions drawn from it}, in The collected papers of Albert Einstein.
  Vol. 2 : The Swiss years: writings, 1900---1909 (Princeton University Press,
  Princeton, New Jersey, 1989)]}.

\bibitem{Darwin1913}
\bibinfo{author}{Darwin, C.~G.}
\newblock \bibinfo{journal}{\bibinfo{title}{On some orbits of an electron}}.
\newblock {\emph{{Philos.~Mag.}}} \textbf{\bibinfo{volume}{25}},
  \bibinfo{pages}{201} (\bibinfo{year}{1913}).

\bibitem{Sommerfeld}
\bibinfo{author}{Sommerfeld, A.}
\newblock \bibinfo{journal}{\bibinfo{title}{Zur {Quantentheorie} der
  {Spektrallinien}}}.
\newblock {\emph{{Ann. Phys. (Berlin)}}}
  \textbf{\bibinfo{volume}{51}}, \bibinfo{pages}{1} (\bibinfo{year}{1916}).

\bibitem{Chakraborty2017}
\bibinfo{author}{Chakraborty, C.} \emph{et~al.}
\newblock \bibinfo{journal}{\bibinfo{title}{Distinguishing kerr naked
  singularities and black holes using the spin precession of a test gyro in
  strong gravitational fields}}.
\newblock {\emph{{Phys. Rev. D}}} \textbf{\bibinfo{volume}{95}},
  \bibinfo{pages}{084024}, \doiprefix\url{10.1103/PhysRevD.95.084024}
  (\bibinfo{year}{2017}).

\bibitem{reldynmagnet}
\bibinfo{author}{Rafelski, J.}, \bibinfo{author}{Formanek, M.} \&
  \bibinfo{author}{Steinmetz, A.}
\newblock \bibinfo{journal}{\bibinfo{title}{Relativistic dynamics of point
  magnetic moment}}.
\newblock {\emph{{Eur.~Phys.~J.~C}}}
  \textbf{\bibinfo{volume}{78}}, \bibinfo{pages}{6} (\bibinfo{year}{2018}).

\bibitem{DiLorenzoprep}
\bibinfo{author}{{Di Lorenzo}, A.}
\newblock \bibinfo{title}{Spacetime accounting for non inertial observers}.
\newblock \bibinfo{note}{In preparation}.

\bibitem{DiLorentz}
\bibinfo{author}{{Di Lorenzo}, A.}
\newblock \bibinfo{title}{A new eigenvalue-based classification of {Lorentz}
  transformations}.
\newblock \bibinfo{note}{To be submitted}.

\bibitem{Synge}
\bibinfo{author}{Synge, J.~L.}
\newblock \emph{\bibinfo{title}{Relativity: The Special Theory}}
  (\bibinfo{publisher}{North-Holland Publishing Company},
  \bibinfo{year}{1956}).

\bibitem{Stapp}
\bibinfo{author}{Stapp, H.~P.}
\newblock \bibinfo{journal}{\bibinfo{title}{Relativistic theory of polarization
  phenomena}}.
\newblock {\emph{{Phys.~Rev.}}} \textbf{\bibinfo{volume}{103}},
  \bibinfo{pages}{425} (\bibinfo{year}{1956}).

\bibitem{Ritus1961}
\bibinfo{author}{Ritus, V.~I.}
\newblock \bibinfo{journal}{\bibinfo{title}{Transformations of the
  inhomogeneous {Lorentz} group and the relativistic kinematics of polarized
  states}}.
\newblock {\emph{{Zh. Eksp. Teor. Fiz.}}}
  \textbf{\bibinfo{volume}{40}}, \bibinfo{pages}{352} (\bibinfo{year}{1961}).
\newblock \bibinfo{note}{[Sov. Phys. JETP \textbf{13} 240 (1961)]}.

\end{thebibliography}

\end{document}